\documentclass[11pt]{article}
\usepackage{geometry}
\usepackage{graphicx}
\usepackage{slashed}
\usepackage{graphicx}
\usepackage{amsmath}
\usepackage{epstopdf}
\usepackage{verbatim}
\usepackage{axodraw}
\usepackage{url}
\usepackage{amsfonts}
\usepackage{layout}
\usepackage{slashed}
\newcommand{\dash}{\text{-}}

\newcommand{\lra}{\leftrightarrow}
\newcommand{\eps}{\epsilon}
\def\a{\alpha}
\def\b{\beta}
\def\g{\gamma}
\def\d{\delta}
\def\m{\mu}
\def\n{\nu}
\newcommand{\da}{\dot{\alpha}}
\newcommand{\db}{\dot{\beta}}
\newcommand{\gd}{\dot{\gamma}}
\newcommand{\dd}{\dot{\delta}}

\def\l{\lambda}
\def\k{\kappa}
\def\th{\theta}
\newcommand{\shk}{r}

\renewcommand{\theequation}{\thesection.\arabic{equation}}

\newcommand{\dak}{\>\!\!\bigr\>}
\newcommand{\dab}{\bigr\<\!\!\<}
\def\dsk{]\!\bigr]}
\newcommand{\dsb}{\bigr[\![}

\newcommand{\dakB}{\hspace{1pt}\>\!\!\bigr\>{\hspace{-2mm}\bullet\hspace{0mm}}}
\newcommand{\dabB}{\,\bigr\<\!\!\<{\hspace{-2.1mm}\bullet\hspace{0mm}}\hspace{1pt}}
\newcommand{\dsbB}{\bigr[\![{\hspace{-2.0mm}\bullet\hspace{0mm}}\hspace{1pt}}
\newcommand{\dskB}{\hspace{1pt}]\!\bigr]{\hspace{-2.1mm}\bullet\hspace{0mm}}}

\newcommand{\cn}{{\cal N}}
\newcommand{\be}{\begin{equation}}
\newcommand{\bea}{\begin{eqnarray}}
\newcommand{\beq}{\begin{equation}}

\newcommand{\ee}{\end{equation}}
\newcommand{\eea}{\end{eqnarray}}
\newcommand{\eeq}{\end{equation}}
\newcommand{\lsim}{\!\mathrel{\hbox{\rlap{\lower.55ex \hbox{$\sim$}} \kern-.34em \raise.4ex \hbox{$<$}}}}
\newcommand{\gsim}{\!\mathrel{\hbox{\rlap{\lower.55ex \hbox{$\sim$}} \kern-.34em \raise.4ex \hbox{$>$}}}}

\newcommand{\reef}[1]{(\ref{#1})}
\newcommand{\<}{\langle}
\renewcommand{\>}{\rangle}
\newcommand{\mhat}{\!\widehat{\,-\,}\!}

\newcommand{\hg}{b}

\newcommand{\example}{\noindent\hangindent 1cm\hangafter0{\em Example.}~~}
\newcommand{\examples}{\noindent\hangindent 1cm\hangafter0{\em Examples.}~~}

\begin{document}

\begin{titlepage}

\begin{flushright}
MCTP-10-46 \\
PUPT-2351 \\
\end{flushright}
\vspace{3.0cm}

\begin{center}
{\Large \bf On-shell constructibility of tree amplitudes\\[1ex] in general field theories}

\vspace{0.2in}
{\bf Timothy Cohen$^a$, Henriette Elvang$^{a,b}$ and Michael Kiermaier$^c$}\\
\vspace{0.2cm}
{\it $^a$Michigan Center for Theoretical Physics (MCTP)\\Department of Physics, University of Michigan, \\ Ann Arbor, MI 48109}\\
\vspace{0.2cm}
{\it $^b$Institute for Advanced Study\\ Princeton, NJ 08540}\\
\vspace{0.2cm}
{\it $^c$Department of Physics, Princeton University \\ Princeton, NJ 08544}\\
\end{center}

\vspace{0.4cm}

\begin{abstract}

We study ``on-shell constructibility'' of tree amplitudes from recursion relations in general 4-dimensional local field theories with any type of particles, both massless and massive.  Our analysis applies to renormalizable as well as non-renormalizable interactions, with or without supersymmetry. We focus on recursion relations that arise from complex deformations of all external momenta. Under certain conditions, these ``all-line shift recursion relations'' imply the MHV vertex expansion. We derive a simple sufficient criterion for the validity of the all-line shift recursion relations. It depends only on the mass dimensions of the coupling constants and on the sum of helicities of the external particles. Our proof is strikingly simple since it just relies on dimensional analysis and little-group transformation properties. In particular, the results demonstrate that all tree amplitudes with $n>4$ external states are constructible in any power-counting renormalizable theory. Aspects of all-line shift constructibility are illustrated in numerous examples, ranging from pure scalar theory and the massless Wess-Zumino model to theories with higher-derivative interactions, gluon-Higgs fusion, and $Z$-boson scattering. We propose a sharp physical interpretation of our constructibility criterion: the all-line shift fails precisely for those classes of $n$-point amplitudes that can receive local contributions from independent gauge-invariant $n$-field operators.
\end{abstract}

\end{titlepage}

\tableofcontents

\newpage

\setcounter{equation}{0}
\section{Introduction}
Amplitudes are ``on-shell constructible" when they can be expressed recursively in terms of lower-point on-shell amplitudes.  Many studies of on-shell constructibility have focused on particular theories, such as massless gauge theories and gravity. In this work, we broaden the scope and ask, quite generally, {\em for which four-dimensional Lorentz-invariant local field theories are the tree amplitudes on-shell constructible?} In an early approach to this question, Benincasa and Cachazo \cite{Benincasa:2007xk} (see also \cite{Schuster:2008nh}) obtained a set of known constraints on interacting theories by requiring 4-point tree amplitudes to be constructible by BCFW recursion \cite{Britto:2004ap,Britto:2005fq}.  Here we take a different route and show that the question of constructibility  has a simple answer in the framework of {\em all-line shift recursion relations}.  We find that in order to determine whether a particular amplitude is
all-line shift
constructible, one needs to know only the external particles of the amplitude and the mass dimensions of the couplings that contribute to it. No
gauge-dependent
analysis of Feynman diagrams or theory-specific properties such as symmetries, full particle content or details of the interactions are needed to establish constructibility in this framework. Our condition for on-shell constructibility has a natural interpretation in terms of the possibility of additional local contributions from gauge-invariant interactions. Such contributions, if present, contain information that cannot be captured by the lower-point amplitudes that enter the recursion relation.

Let us briefly review the method of using complex momentum shifts \cite{Britto:2005fq} to derive on-shell recursion relations for tree amplitudes. One applies a complex shift to two or more of the external momenta $p_i \to p_i + z \, q_i$. The shift is arranged such that it preserves momentum conservation and leaves all external momenta on-shell.  The on-shell amplitude $\hat{A}_n(z)$ is then a rational function of the complex parameter $z$, and at tree level it has only simple poles.  If $\hat{A}_n(z) \to 0$ as $z \to \infty$, Cauchy's theorem for the function $\hat{A}_n(z)/z$ expresses the unshifted amplitude $A_n \!=\! \hat{A}_n\bigr|_{z=0}$ (the residue at $z=0$) as a sum of all the other residues.  At each pole, an internal state goes on-shell and factorization gives the residue in terms of lower-point on-shell amplitudes.  Demonstrating that $\hat{A}_n(z) \to 0$ as $z \to \infty$ is at the heart of proving the validity of the corresponding recursion relation.

In this paper, we work with all-line shift recursion relations, which arise from  shifts that deform every external momentum of an amplitude.  An all-line shifts acts democratically on the external lines; it deforms the momentum of each external state in the same way, independent of quantum numbers and particle type. This property makes the all-line shift a universal tool that applies to very
general local field theories.
 Another
important aspect of  all-line shifts is that they can be used \cite{Elvang:2008vz} to derive and prove the validity of the ``MHV vertex expansion" of CSW \cite{Cachazo:2004by} in (super) Yang-Mills theory, (S)YM.
We discuss when the MHV vertex expansion results from all-line shifts for general theories, but more broadly we study the all-line shift recursion relations in their own right.

In a four-dimensional theory with only massless particles, the all-line shift can be implemented as an \emph{anti-holomorphic shift} of the form $|i] \to |i] + z\, w_i \, |X]$. Here $|X]$ is a reference spinor and the $w_i$'s are complex numbers chosen such that momentum conservation is satisfied.  We show that the worst-possible large-$z$ behavior of an amplitude is governed by the simple formula:
\bea\label{falloff1}
   \hat{A}_n(z) \to z^{s}\,~\text{~~~as}~~~z \to \infty\,,~\quad\text{with }\quad 2s =  4-n -c +H \,~~~~~~ \mbox{(anti-holomorphic shift)}\, .
\eea
Here, $n$ is the number of external states, $c$ is the mass dimension of the product of couplings in this amplitude,%
\footnote{If more than one product of couplings appears, $c$ is the smallest mass dimension;
see section \ref{AllLineShifts}.}
and $H=\sum_i h_i$ is the sum of helicities of the external states of $A_n$ (all outgoing).  The proof is
strikingly simple,
relying only on dimensional analysis and the little-group transformation properties of amplitudes.  The result \reef{falloff1} applies to Lorentz-invariant local theories with massless particles of any spin $\le 2$ and with couplings of any mass dimension; it is valid for both renormalizable and non-renormalizable theories.

Armed with the result \reef{falloff1}, one can easily determine whether all-line shift recursion relations are valid for a given model. As an example, consider
(S)YM
theory: the coupling is dimensionless, so $c=0$. An N$^k$MHV gluon amplitude has $k+2$ negative helicity states and $n-k-2$ positive helicity states, and hence $H=n-2k-4$; this holds for any N$^k$MHV amplitude of the theory.  Inserting these values for $c$ and $H$ into \reef{falloff1}, we immediately find $\hat{A}^\text{\tiny N$^k$MHV}_n(z) \to z^{-k}$ as $z \to \infty$, and therefore the anti-holomorphic all-line shift recursion relations are valid for all amplitudes beyond the MHV level.

Suppose we consider instead a \emph{holomorphic} all-line shift, $|i\> \to |i\> + z\, \tilde{w}_i \, |X\>$.
It
gives a large-$z$ behavior:
\bea\label{falloff1b}
   \hat{A}_n(z) \to z^{a}\,~\text{~~~as}~~~z \to \infty\,,~\quad\text{with }\quad 2a =  4-n -c -H \,~~~~~~ \mbox{(holomorphic shift)}\, .
\eea
This can be combined with \reef{falloff1} to give $a+s \leq 4-n-c$. In a  (power-counting) renormalizable theory, the amplitudes have $c\ge 0$, so $a+s \leq 4-n$. Thus for $n>4$, either $a$ or $s$ (or both)
 will be negative, and this means that there exists a shift such that $\hat{A}_n(z) \to 0$ for large $z$. Therefore,
\emph{in power-counting renormalizable theories, all amplitudes with $n>4$ external states are on-shell constructible using all-line shift recursion relations.} This is a very general result derived by simple means. Applying it to (S)YM theory, we note that N$^k$MHV amplitudes can be computed with anti-holomorphic shifts for $k>0$ while the $k=0$ MHV amplitudes can be computed with holomorphic shifts for  $n>4$.

So far we have discussed theories with massless particles.  To study massive particles,
we first need
a proper generalization of the \mbox{(anti-)holomorphic} all-line shift. It turns out that this generalization is essentially unique because there are strong constraints on obtaining a consistent definition of all-line shifts with massive external particles. Secondly, we need to determine the large-$z$ behavior of amplitudes under such a massive all-line shift. Giving masses to particles  should not change the large-$z$ behavior of amplitudes, because $z\to \infty$ is a UV limit; as the momenta are taken very large, the masses become irrelevant. This intuition is correct, but the practical implementation requires a little more care. We
use a massive spinor-helicity formalism,%
\footnote{Appendix \ref{sec:SpinorHelicityFormalism} summarizes our conventions for the massive spinor-helicity formalism.}
based on the work of Dittmaier \cite{Dittmaier:1998nn}, in which the massive momenta are decomposed along a reference null direction $q$.
It allows us to assign a ``$q$-helicity'' $\tilde{h}_i$ to each particle.
This
is nothing but a way to label the particles in terms of eigenstates of
a
$q$-dependent helicity operator. In the massless limit, $\tilde{h}_i$ is just the ordinary frame-independent helicity $h_i$. We prove that the large-$z$ behavior \reef{falloff1} holds for models with massive states with the sum of helicities $H$ replaced by the sum of $q$-helicities $\tilde{H}= \sum_i \tilde{h}_i $. The proof is a little more elaborate than in the massless case. For example, one needs to account for the longitudinal polarizations of massive vectors; we use the Goldstone boson equivalence theorem to treat these states.%
\footnote{The equivalence theorem was previously used in the context of large-$z$ behavior in \cite{Boels:2009bv}.}
Also, there are amplitudes that are non-vanishing in the massive case only, and they have to be studied separately. As in the massless case, the large-$z$ analysis shows that all amplitudes with $n>4$ legs in power-counting renormalizable theories are constructible using all-line shift recursion relations.

There is
a
natural interpretation of the sufficient condition for all-line shift constructibility, \mbox{$4-n-c-|H|<0$}.
 The point is the following: if a recursion relation is valid for an $n$-point amplitude, all information  about this process is already encoded in the amplitudes with less than $n$ external legs. In particular,  there can be no independent information provided by the $n$-point contact term interactions in the theory. If such interactions are present in the Lagrangian, they must be {\em dependent interactions} that are either completely determined from lower-point interactions by gauge-invariance, or that can be absorbed into the lower-point interactions by a field redefinition. A familiar example is the 4-point interaction in Yang-Mills theory: a gauge can always be chosen to make its contribution to the on-shell 4-point gluon amplitude vanish. The 4-vertex is present to ensure off-shell gauge invariance of the Yang-Mills Lagrangian, but it plays no role for the physical 4-point amplitude. Our examples indicate that if an interaction $Y$ is required to preserve off-shell gauge invariance in the presence of some lower-point interactions $X$, then only the input of $X$ is needed
to construct the tree amplitudes.

An {\em  independent interaction}, however, requires
separate input. For example, the information contained in the scalar interaction $\l \phi^4$ cannot be obtained from lower-point on-shell amplitudes. This is indeed the reason why constructibility only starts at 5-points in general renormalizable theories. As we will explain, this analysis applies much more generally, for example in gauge theory with interactions from higher-dimensional operators such as $ D^{2q}F^m$. We cannot expect the $n$-point gluon matrix elements of this operator to be on-shell constructible for
$n\!=\!m\!+\!1,$ $m\!+\!2,\ldots,m\!+\!q$, because they could receive
local
contributions from
 gauge-invariant operators of the form $D^{2q-2}\!F^{m+1}$, $D^{2q-4}\!F^{m+2}$, $\ldots$, $F^{m+q}$, which all have the same coupling dimension. Indeed, all-line shift constructibility for the gluon matrix elements of $ D^{2q}F^m$ fails for $n\leq m+q$. We use this and other examples to give a sharp physical interpretation of the all-line shift constructibility bound.

What about interactions related by symmetries, for example supersymmetry? We propose that amplitudes with such dependent interactions are on-shell constructible only when the recursion relations incorporate the relevant symmetry. All on-shell recursion relations build in gauge invariance, but  supersymmetry,
for example,
requires one to work with super-shifts of superamplitudes. This is indeed done in super-BCFW \cite{ArkaniHamed:2008gz,Brandhuber:2008pf}  and the supersymmetric version of (anti-)holomorphic shifts \cite{Kiermaier:2009yu}. We discuss
the interpretation of the constructibility bound in
more
detail in section \ref{sec:DiscussionOfConstructibility}.

Let us compare and contrast our work with previous analyses. Complex shifts were first introduced by Britto-Cachazo-Feng-Witten \cite{Britto:2005fq} as a method for deriving on-shell recursion relations in Yang-Mills theory. Proofs of the validity of BCFW recursion relations have typically required detailed
analyses
of the large-$z$ behavior of individual theory-specific `dangerous' Feynman diagrams \cite{Benincasa:2007qj,ArkaniHamed:2008yf,Cheung:2008dn}. The light-cone gauge approach introduced by Arkani-Hamed and Kaplan \cite{ArkaniHamed:2008yf} was generalized by Cheung \cite{Cheung:2008dn} to show BCFW-constructibility in a large class of 2-derivative gauge and gravity  theories without higher-point gauge-invariant operators such as $\phi^m F^2$. The analysis \cite{Cheung:2008dn} applies to amplitudes with a gauge boson among the external states. Our methods here do not require information about specific Feynman diagrams or gauge choices, but are manifestly
gauge-invariant.
The condition for constructibility allows any type of local interactions, with any number of derivatives, including for example $\phi^m F^2$ and $D^{2q} F^m$. A physically relevant example of such a
 higher-dimensional
operator is the gluon-Higgs effective operator $h \, {\rm tr} F^2$.

The all-line shift is inspired by the anti-holomorphic Risager shift \cite{Risager:2005vk}, which acts only on the $k\!+\!2$
negative helicity gluon lines in N$^k$MHV gluons amplitudes. It was shown in \cite{Risager:2005vk,BjerrumBohr:2005jr} that iterative use of the resulting recursion relations give the CSW expansion for gluon amplitudes. This expansion allows one to express any on-shell gluon N$^k$MHV amplitude in terms of $k\!+\!1$ MHV gluon amplitudes, which are given by the compact Parke-Taylor formula for any number of external legs. For this reason, the CSW expansion is also known as the MHV vertex expansion. The MHV vertex expansion was extended to all tree amplitudes of $\cn=4$ super Yang-Mills theory; this was proven in \cite{Elvang:2008vz} using all-line shift recursion relations, following earlier work \cite{Nair:1988bq,Georgiou:2004by, Bianchi:2008pu, Elvang:2008na}. The result~(\ref{falloff1}) for the large-$z$ falloff in general theories gives, in particular, an alternative, much simpler derivation of the validity of all-line shift recursion relations in $\cn=4$ SYM.

{\em The paper is organized as follows.}  We introduce the all-line shifts and study their large-$z$
behavior
in section \ref{AllLineShifts}. In section \ref{AllToMHV} we discuss a sufficient set of criteria for the all-line shift recursion relations to produce an
MHV vertex expansion.
Section \ref{s:examples} is dedicated to a variety of examples that illustrate various properties of the massless all-line shift recursion relations.  The generalization to massive external states is given in section \ref{sec:MassiveParticles}: it includes a proof of our general result \reef{falloff1} for massive particles  and some examples.  Finally, in section \ref{sec:DiscussionOfConstructibility} we give a physical interpretation of why some amplitudes are not constructible via all-line shift recursion relations. Appendix \ref{sec:allline2CSW}
outlines the derivation of
 the MHV vertex expansion from the
all-line shift recursion relations,
 and appendix \ref{sec:SpinorHelicityFormalism} summarizes our conventions for the massive spinor-helicity formalism.

\setcounter{equation}{0}
\section{All-line shift recursion relations}\label{AllLineShifts}
In this section, we introduce the all-line shift and derive the explicit formula \reef{falloff1} for the large-$z$ behavior of amplitudes\footnote{Henceforth `amplitude' means tree-level helicity amplitude unless otherwise stated.} in any 4-dimensional Lorentz-invariant local field theory of massless particles. It is useful to first introduce three integers --- $c$, $s$, and $a$ --- to characterize the
amplitudes.

\vspace{2mm}
\noindent {\bf Coupling dimension ($c$):} \\
Let $c_i$ be the mass dimensions of the couplings $g_i$ of the interactions in the theory.
Each term $T$ in a given amplitude
involves a certain product of couplings,
$g_T=\Pi g_i$, with total mass dimension $c_T = \sum c_i$. For the sake of simplicity, let us  assume in the following that all
terms in a given amplitude have the same coupling $g_T=g$ of mass dimension $c_T=c$. We will generalize this towards the end of the section.

\vspace{2.5mm}
\examples
The gluon self-interaction in \emph{Yang-Mills theory} has a dimensionless coupling, so any amplitude in this theory has $c=0$. In perturbative \emph{Einstein gravity}, all interactions involve two derivatives and take the schematic form $\kappa^{n-2} \partial^2 h^n$. The coupling $\kappa$ has mass dimension $-1$, and one can show that any $n$-point amplitude has $c=2-n$.

\vspace{2.5mm}
\noindent Table \ref{tab:cValues} lists values of $c$ for amplitudes in various theories studied in the examples of section \ref{s:examples}.

\vspace{2mm}
\noindent {\bf Angle and square brackets ($a$ and $s$):} \\
In spinor-helicity formalism, an on-shell tree amplitude with only massless particles is a rational function of angle and square brackets. Collecting all contributions into a single term with a common denominator, we write an $n$-point on-shell amplitude $A_n$ schematically as
\bea
\label{Astructure}
A_n ~=~g ~ \frac{ \sum \,\<..\>^{a_n}\,[\,..\,]^{s_n}}{\sum\,\<..\>^{a_d}\,[\,..\,]^{s_d}} \, .
\eea
The numerator and denominator contain sums of products of angle and square brackets. As we shall see below, little-group transformation properties and dimensional analysis require that each term in the numerator has the same number of angle brackets and the same number of square brackets (and similarly for the denominator). Therefore it is meaningful to introduce $a$ and $s$ as the difference between the number of angle/square brackets in the numerator and denominator:
\begin{equation}
\begin{split}\label{defas}
  a &\equiv
  (\text{\# of $\<..\>$'s in  numerator}) - (\text{\# of $\<..\>$'s in  denominator})\,, \\
  s &\equiv
  (\text{\# of $[\,..\,]$'s in  numerator}) - (\text{\# of $[\,..\,]$'s in  denominator})\,.
\end{split}
\end{equation}
The integers $a$ and $s$ are useful for characterizing the tree amplitude.

\vspace{2.5mm}
\examples
The Parke-Taylor MHV gluon amplitude
\bea\label{eq:ParkeTaylor}
  \< --+\dots + \>
   =
  \frac{\<12\>^4}{\<12\>\<23\>\cdots\<n1\>}
  ~~~~~~~~~\text{has}~~~~~
  \biggl\{
     \begin{array}{l}
     a = 4-n\\
     s = 0
     \end{array}
  \,.
\eea
The NMHV gluon amplitude $\<---+++\>$ can be written as
\bea
  \<---+++\>=\frac{\<1|2+3|4]^3}{s_{234} [23] [34] \<56\> \<6 1\> \< 5| 3+4 |2]}
  +
  \frac{\<3|4+5|6]^3}{s_{612} [61] [12] \<34\> \<45\> \< 5| 3+4 |2]} \, .~
  \label{3m3p}
\eea
Each element $\<i|j+k|l] = \<ij\>[jl]+ \<ik\>[kl]$ or $s_{ijk} = -\<ij\>[ij]- \<ik\>[ik]-\<jk\>[jk]$ contains one power of angle brackets and one of square brackets. Thus both terms in \reef{3m3p}  have $a = 3 - 4=-1$ and $s= 3-4=-1$. In fact,
it can be shown that
\begin{equation}\label{YM}
    \text{(super) Yang-Mills:}~~\qquad a = 4-n+k\,, \qquad s = -k
\end{equation}
for N$^k$MHV amplitudes in pure (super) Yang-Mills theory.

\begin{table}[t!]
\begin{center}
\begin{tabular}{|c||c|c|c|c|c|c|c|c|}
\hline
 theory
 & $\phi^3$ & $\phi^4$ & (S)YM &
 (s)gravity &  $\phi^m,$~{\footnotesize $m>4$} & $z F^2$ & $D^{2q}F^m$ & $R^m$  \\
\hline
 $c$   & $n\!-\!2$  & 0& 0     & $2\!-\!n$   &
  $-x(m-4)$
 & $-x$    &
 $\!\!-x(2m\!+\!2q\!-\!4)\!$
 &
 $\!2\!-\!n\!-2x(m\!-\!1)\!$
 \\
\hline
\end{tabular}
\end{center}
\vspace{-3mm}
\caption{\small Values of $c$ for amplitudes in various models. $n$ is the number of external states, and $x$ denotes the number of insertions of the given higher-dimensional operator.}
\label{tab:cValues}
\end{table}

\vspace{2.5mm}
\noindent Next, we explain how the couplings and the external states restrict $a$ and $s$.

\vspace{2mm}
\noindent {\bf Dimensional analysis ($a+s$):} \\
The mass dimension of an $n$-point amplitude in four dimensions is $4\!-\!n$. Angle and square brackets have mass dimension 1, so with a coupling $g$ of mass dimension $c$ in \reef{Astructure} we must have
\bea
   \label{asc}
   \boxed{\phantom{\Biggl(}
   a+s = 4-n -c\,.~~}
\eea

\vspace{2.5mm}
\examples
Amplitudes in Yang-Mills theory have $c=0$, so \reef{asc} gives $a+s=4-n$. This is consistent with~(\ref{YM}). In gravity, $c=2-n$ and hence
graviton
tree amplitudes have $a+s=2$, independently of $n$.

\vspace{2mm}
\noindent {\bf Little-group scaling ($a-s$):} \\
Amplitudes scale homogeneously under ``little-group scalings''. Specifically, for each external state $i$ with helicity $h_i$ we have
\bea\label{little}
  |i\> \to t_i |i\>\, ,
  ~~~~~
  |i] \to t_i^{-1} |i]
  ~~~~\implies~~~~
  A_n \to t_i^{-2h_i} A_n \, .
\eea
If all spinors $|i\>$ and $|i]$ transform with the same parameter $t$,
then
the  amplitude scales as
\mbox{$A_n \to t^{-2\sum_i h_i} A_n$.}
At the same time, the structure \reef{Astructure} implies that $A_n \to t^{2(a-s)} A_n$. We conclude that
\bea
  \label{aMs}
     \boxed{\phantom{\Biggl(}
	a-s = - \sum_i h_i \, .~~}
\eea
Combining \reef{asc} and \reef{aMs} gives
\bea
  \label{thes}
   2 s = 4-n -c +\sum_i h_i\, ,~~~~~~~~~~
   2 a = 4-n -c - \sum_i h_i\,.
\eea
Thus $a$ and $s$ are completely determined by dimensional analysis and little-group scaling; that is why $a$ and $s$ as given in~(\ref{defas}) are well-defined quantities and cannot differ from term to term in the amplitude.

\vspace{2mm}
\noindent {\bf Anti-holomorphic all-line shifts:} \\
We focus on a momentum-conserving anti-holomorphic shift of all the external lines:
\bea\label{allline}
   |i]~\to~|\hat{i}] = |i] + z\,w_i\, |X] \, ,
   ~~~~~\text{with}~~~~\sum_i w_i |i\> = 0 \, .
\eea
For generic external momenta and generic $|X]$, all square brackets $[ij]$ shift linearly in $z$, while the angle brackets remain unshifted. The preceding analysis shows that
\bea\label{falloff}
\boxed{\phantom{\Biggl(}
   \hat{A}_n(z) \to z^s\,~\text{(or better)~~~as}~~~z \to \infty\,,~\quad\text{with }\quad
   2s = 4-n -c +\sum_i h_i \, .
   ~}~
\eea
The condition $s<0$ is sufficient to ensure $\hat{A}_n(z) \to 0$ for large $z$. This means that there is no pole at infinity, and in the usual way
\cite{Britto:2005fq}
Cauchy's theorem then gives a valid recursion relation for the amplitude:\footnote{
For a holomorphic all-line shift,  $|\hat{i}\> = |i\> +  \tilde{w}_i\, |X\>$, the integer $a$ controls the large-$z$ behavior. Hence $a<0$ is a sufficient condition for the validity of \emph{holomorphic} all-line shift recursion relations.
}
\bea
    0 = \oint_\mathcal{C} \frac{A_n(z)}{z}
    ~~~~\implies~~~
    A_n = \hat{A}_n(0) = \sum_I \hat{A}_{L}(z_I) \frac{1}{P_I^2}  \hat{A}_{R}(z_I)
    \, .
    \label{arecrel}
\eea
The sum is over all tree diagrams with subamplitudes $\hat{A}_{L}$ and $\hat{A}_{R}$ evaluated at shifted momenta with $z=z_I$ such that the internal line $\hat{P}_I$ is on-shell. In certain cases (to be discussed in the next section), the all-line shift recursion relation is equivalent to the MHV vertex expansion.

\vspace{2mm}
\noindent {\bf General couplings ($c$ again):} \\
In the above discussion we assumed that the product of couplings for each
term $T$
has  the same mass dimension $c=c_T$. If this is not the case, our discussion above implies that the \emph{worst} falloff for large $z$ arises from
terms
with the \emph{smallest} value of $c_T$. The $c$ that appears in \reef{falloff} is therefore defined to be the \emph{smallest} value $c_T$ of any
term in
the amplitude.

\vspace{2.5mm}
\example
For purely illustrative purposes, let us consider scalars with cubic and quartic interactions, schematically $\mu \phi^3$ and $\lambda \phi^4$. A 4-point amplitude can have contributions from pole diagrams with two cubic vertices and from 4-point contact terms. These diagrams have couplings $g_\text{pole} = \mu^2$ and $g_\text{contact} = \lambda$, respectively,
with mass dimensions $c_\text{pole} = 2$ and $c_\text{contact} = 0$.  Under the all-line shift, the pole diagrams go as $1/z$ for large $z$, while the contact terms are unshifted. Hence the
shifted
4-point
amplitude behaves as $z^0$ for large $z$. Indeed, this is the behavior determined by \reef{falloff} when $c$ is taken to the lowest mass dimension of the couplings $g_\text{pole}$ and $g_\text{contact}$, \emph{i.e.,}~$c=c_\text{contact}=0$.

\vspace{2mm}
\noindent {\bf Power-counting renormalizable theories:}\\
Consider a theory of massless particles whose coupling constants have either vanishing or positive mass dimension; these are ``power-counting renormalizable'' theories. In such a theory $c\geq 0$, and consequently we have
\begin{equation}
    s+a\leq 4-n  \qquad\text{(power-counting renormalizable theory)}\,.
\end{equation}
We conclude that amplitudes with $n>4$ have either $s<0$ or $a<0$ (or both) and therefore vanish either under an anti-holomorphic or a holomorphic all-line shift (or both). {\em In power-counting renormalizable theories, amplitudes with $n>4$ external lines are always constructible from an all-line shift recursion relation.} The derivation made heavy use of the little-group properties appropriate for massless particles, but this result generalizes to the massive case (see section~\ref{sec:MassiveParticles}).

\setcounter{equation}{0}
\section{From all-line shifts to the MHV vertex expansion}
\label{AllToMHV}
We have derived
the
simple sufficient condition
\reef{falloff}
for the validity of the all-line shift recursion relations.
In this section we discuss when the all-line shift recursion relations can be applied iteratively to yield the MHV vertex expansion
\cite{Cachazo:2004by}.

Let us begin with a brief review of the MHV vertex expansion in Yang-Mills theory. The MHV vertex expansion expresses a tree gluon amplitude as a sum of ``MHV vertex diagrams". At the N$^k$MHV level, each diagram contains $k+1$ MHV vertices and $k$ internal lines, for example
\begin{equation}\label{eq:SchematicN3MHV}
    A^{\text{N$^3$MHV}}~=~ \sum
    \parbox[c]{5.5cm}{\includegraphics[width=5.5cm]{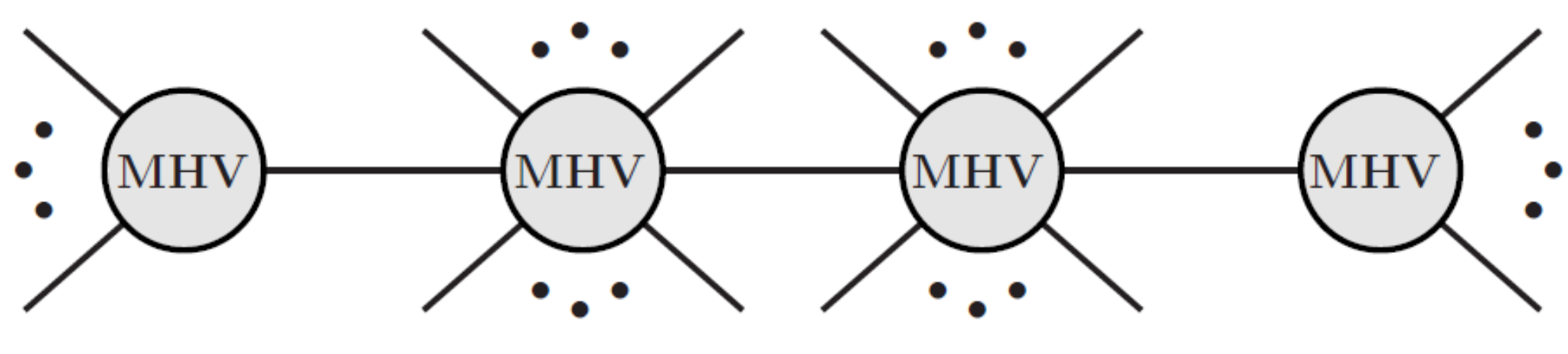}}
    ~+~\sum \parbox[c]{4.5cm}{\includegraphics[width=4.5cm]{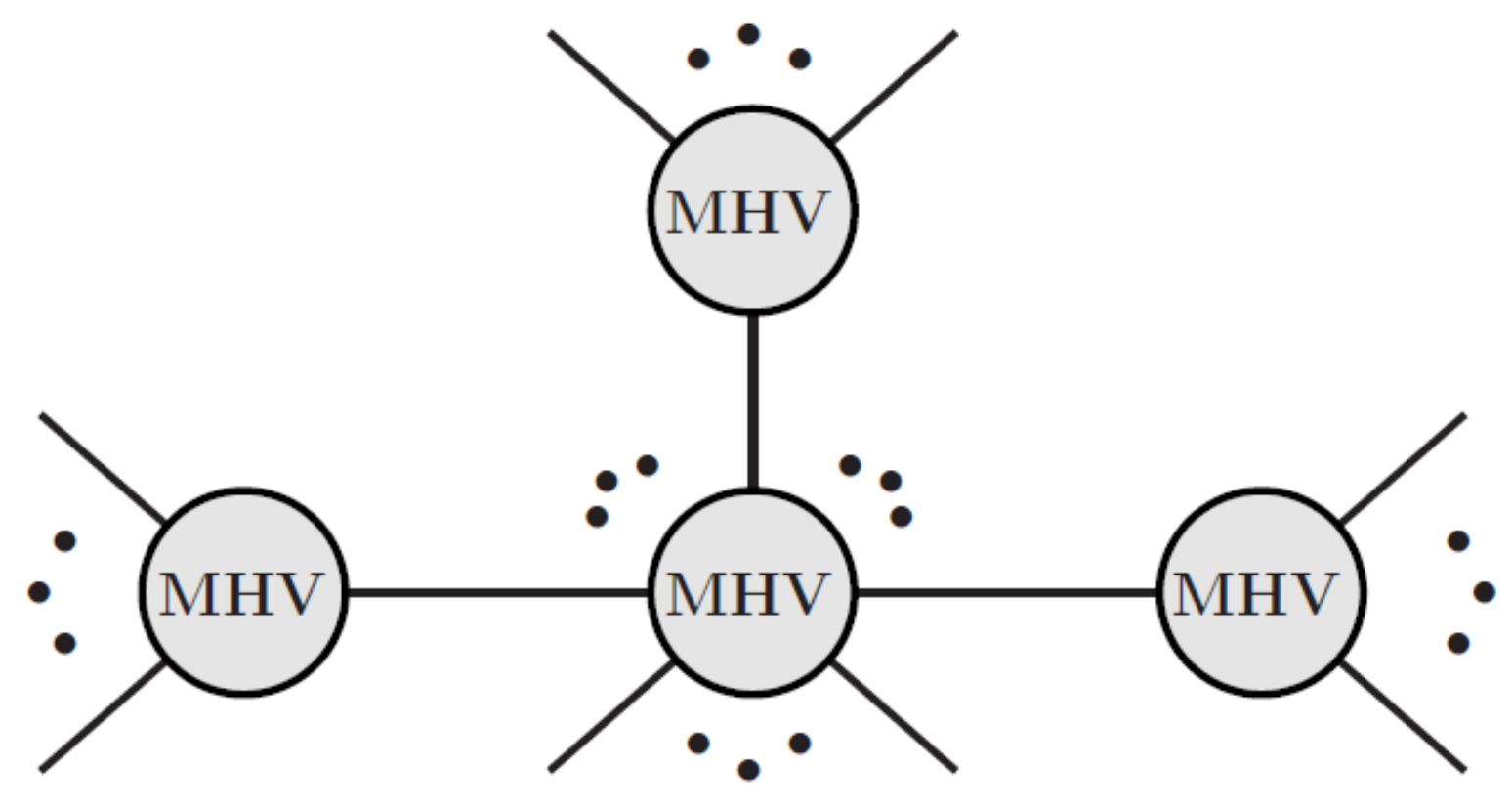}}\,.
\end{equation}
The value of each diagram is simply the product of the $k+1$ on-shell MHV subamplitudes and the $k$ scalar propagators $1/P_I^2$.
For internal
momenta
$P_I$,
one defines angle spinors $|P_I\>$ using the CSW prescription
\begin{equation}\label{CSWpres}
    |P_I\> \equiv P_I|X]\,,
\end{equation}
where $|X]$ is an arbitrary reference spinor. Square spinors $|P_I]$ are not needed because the MHV gluon amplitudes depend only on angle spinors (see \reef{eq:ParkeTaylor}). The sum of all MHV vertex diagrams is independent of $|X]$.

The construction of the MHV vertex diagrams from the diagrams of the all-line shift recursion relations is reviewed in appendix \ref{sec:allline2CSW} in the context of $\cn=4$ SYM. \emph{The essential properties required in this derivation are:}

\begin{enumerate}
  \item A \emph{classification} of amplitudes into N$^k$MHV sectors is needed; or at least a characterization of MHV vs.~non-MHV. This property is guaranteed in a supersymmetrizable theory, as we discuss below.
  \item \emph{All non-MHV  amplitudes must admit a valid anti-holomorphic all-line shift.} By \reef{falloff}, a sufficient condition is that $s<0$ for all non-MHV amplitudes.
  \item \emph{MHV amplitudes are invariant under the anti-holomorphic all-line shift.} A sufficient condition is that the MHV amplitudes depend only on angle brackets.%
\footnote{For the MHV vertex expansion to be useful,
one needs
to be able to construct the tower of MHV amplitudes, for example via BCFW.}
  \item \emph{No anti-MHV 3-point subamplitudes} are allowed in the all-line shift recursion relations. This is ensured by kinematics of the anti-holomorphic shift if the  anti-MHV 3-point amplitudes of the theory vanish as $[12],[23],[31]\to0$.
\end{enumerate}

If these four conditions are satisfied, the all-line shift recursion relations become equivalent to  the MHV vertex expansion. We will see several examples of this in section \ref{s:examples}. It is important to emphasize that whether or not the properties 1-4 hold, the validity of the all-line shift recursion relations relies only on the large-$z$ falloff in \reef{falloff}, \emph{i.e.,}~$s<0$.

\vspace{3mm}
\noindent {\bf N$^k$MHV classification:}\\
In preparation for the examples, we outline the general  N$^k$MHV classification of amplitudes. In a four-dimensional $\cn=1$ supersymmetric theory, states and their annihilation operators can be classified as `$\alpha$' or `$\beta$'  depending on whether they are annihilated by $Q$ or $\tilde Q$ \cite{Grisaru:1976vm,Grisaru:1977px}:
\begin{equation}
   \label{abstates}
    [Q,\alpha]=0\ ,\qquad
    [\tilde{Q},\alpha] = \<\epsilon\, p\>\,\beta\,,
    \qquad
    \qquad
    [\tilde{Q},\beta]=0\,, \qquad
    [Q,\beta] = [\epsilon\, p]\,\alpha\,.
\end{equation}
In $\cn =1$ SYM theory, for example, a negative helicity gluon is an $\alpha$-state while a  negative helicity gluino is a $\beta$-state. The positive helicity gluon and gluinos are $\beta$- and $\alpha$-states, respectively.

All amplitudes with $n\geq4$ external states must include at least two $\alpha$-states and two $\beta$-states; if they have
fewer $\a$'s or $\b$'s, the SUSY Ward identities force them to vanish.\footnote{This holds
when all external states are massless. In supersymmetric theories with massive particles, amplitudes with only one $\alpha$- or only one $\beta$-state can also be non-vanishing.} Amplitudes with $m$ $\alpha$-states and $(n-m)$ $\beta$-states are N$^{(m-2)}$MHV.

In theories with extended supersymmetry, each set of supercharges has  an associated $\alpha$,\,$\beta$ classification. For example, in $\cn =4$ SYM theory a negative helicity gluon is an $\alpha$-state of all four supercharges, but a negative helicity gluino is an $\alpha$-state of three  supercharges and a $\beta$-operator of the fourth supercharge. When a theory is invariant under a global $R$-symmetry relating all supercharges, the N$^k$MHV classification is defined as in an $\cn=1$ theory. However, if the $R$-symmetry is broken, a classification level $k_a$ is needed for each unrelated set of supercharges $Q^a$. For example, two integers $k$ and $\tilde{k}$ are required to classify closed string tree amplitudes with massless external states in four dimension, since  only a $SU(4)\times SU(4)$ subgroup of the  $SU(8)$ $R$-symmetry is preserved \cite{Elvang:2010kc}. We will not encounter multiple classification levels for the examples in the following sections.

\setcounter{equation}{0}
\section{Examples}
\label{s:examples}

In this section, we illustrate various aspects of the all-line shift recursion relations and the MHV vertex expansion with several examples. We begin with a discussion of the three simplest theories: $\phi^4$-theory, pure (super) Yang-Mills theory and pure supergravity. Then we turn to more interesting examples, namely the Wess-Zumino model, gluon-Higgs fusion, and theories with higher-derivative operators such as $F^m$ in gauge theory. We end this section with an all-line shift proof of the well-known formula for the rational all-minus 1-loop amplitude in QCD.

\subsection{The simplest theories}
\label{s:simplestEx}

\noindent {\bf $\phi^4$-theory}
\label{sec:phiFourth}\\[1mm]
Let us apply the analysis of sections \ref{AllLineShifts} and \ref{AllToMHV} to the simplest theory: $\lambda \, \phi^4$ theory. The coupling $\lambda$ is dimensionless, so $c=0$, and scalars have $h_i=0$. With this input, the simple criteria \reef{falloff} immediately shows that the $2m$-point scalar amplitudes have a $1/z^{m-2}$ falloff under holomorphic as well as anti-holomorphic all-line shifts. Hence the all-line shift recursion relations are valid for $m>2$, and the theory is on-shell constructible%
\footnote{It is often stated that  $\phi^4$-theory is not tree-level constructible because BCFW fails. This has been `repaired' in the literature either by introducing auxiliary fields to resolve the 4-point interaction into pole diagrams \cite{Benincasa:2007xk} or by reconstructing the pole at infinity \cite{Feng:2009ei}.}
at tree level with a single input amplitude, namely the 4-point amplitude $\< \phi \phi \phi \phi \> = \lambda$.

It is clear that $\lambda \, \phi^4$-theory satisfies all the criteria in section \ref{AllLineShifts} for a valid MHV vertex expansion. The MHV sector consists of the constant 4-point amplitude $\< \phi \phi \phi \phi \>$ only. Since this amplitude is trivially on-shell, the MHV vertex expansion is identical to the Feynman diagram expansion.\footnote{This in fact generalizes to massive $\phi^4$ theory, where the all-line shift recursion relations for massive particles that we introduce in section~\ref{sec:MassiveParticles} also precisely reproduce the Feynman diagram expansion. However,  in section~\ref{sec:MassiveParticles} we focus on other, less trivial, examples of the massive all-line shift recursion relations.}

\vspace{3mm}
\noindent {\bf Pure (super) Yang-Mills theory}\\[1mm]
We have already discussed pure (super) Yang-Mills theory in sections \ref{AllLineShifts} and \ref{AllToMHV}, so let us be brief. The N$^k$MHV amplitudes of (S)YM theory fall off as $1/z^k$ for large $z$ and the all-line shift recursion relations imply the MHV vertex expansion. MHV amplitudes ($k=0$) cannot be calculated from anti-holomorphic all-line shift recursion relations, but as discussed in the Introduction, they can be computed with holomorphic all-line shift recursion relations for $n>4$.

\vspace{3mm}
\noindent {\bf Pure supergravity}\\[1mm]
In (super)gravity, the coupling $\kappa$ has dimension $-1$, and as argued in section \ref{AllLineShifts} this means that $c = 2\!-\!n$ for $n$-point tree amplitudes. For an N$^k$MHV amplitude the sum of helicities is $\sum_i h_i = -2(k\!+\!2) + 2(n\!-\!k\!-\!2) = 2n\!-\!4k\!-\!8$, so $s = n\!-\!3\!-\!2k$. By \reef{falloff}, we then have
\bea
\label{grav}
 \hat{M}_n(z) \sim
  z^{n-3 -2k}~~~\text{as}~~~z \to \infty
\eea
under an all-line shift. This is exactly the behavior expected from the KLT relations, which in field theory take the form $M_n = \sum s_{ij}^{n-3} A_n^2$; here $M_n$ and $A_n$ are the tree-level gravity and color-ordered gauge theory amplitudes, and $s_{ij}^{n-3}$ is a product of $n\!-\!3$ Mandelstam variables.

The large-$z$ behavior \reef{grav} shows that the all-line shift recursion relations are never valid for all amplitudes of a given number of external particles $n$. In particular, amplitudes whose sum of helicities $H$ lies in the range $-2\leq H\leq 2$ cannot be constructed from (holomorphic or antiholomorphic) all-line shift recursion relations for any $n$; these amplitudes are thus required as independent input for the all-line shift recursion relations. Shifts specialized to the external particles, however, such as the all-minus shifts studied in \cite{BjerrumBohr:2005jr,Bianchi:2008pu}, yield a certain shift-dependent MHV vertex expansion for pure-graviton amplitudes that works up to a certain number of external legs; in the NMHV sector, it applies for $n<12$ \cite{Bianchi:2008pu}.

\subsection{The Wess-Zumino model}
\label{sec:WessZumino}
\begin{figure}[t]
\begin{center}
\begin{tabular}{r|c| c}
  &$\alpha$-states & $\beta$-states\\
  \hline$\phantom{\Bigl(}\!\!\!\!\!\!\!\!$
  &~~$f_a\qquad ~~\bar{z}_a$ & $\bar{f}_a ~\qquad z_a$ \\
  \hline$\phantom{\Bigl(}\!\!\!\!\!\!\!\!$
  $U(1)_R$&$-1/3\quad -2/3$ & $1/3\quad~ 2/3$ \\
  \hline$\phantom{\Bigl(}\!\!\!\!\!\!\!\!$
  helicity&-1/2\qquad ~~0~ & 1/2\qquad 0~
\end{tabular}
\hskip6mm
\parbox[c]{7.3cm}{\includegraphics[width=7.3cm]{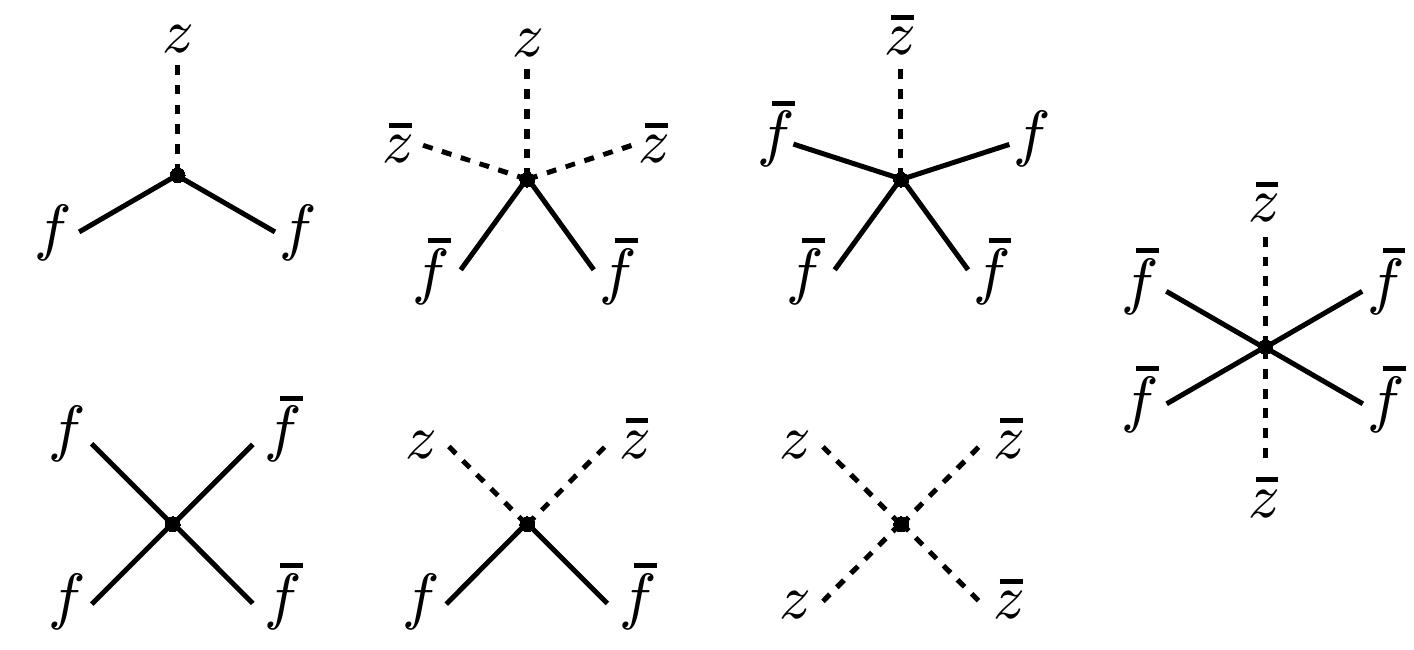}}
\end{center}
\vskip.3cm
\centerline{(a)~~~~\,\hskip7cm ~~(b)}
\caption{\small The particle content (a) and the seven MHV vertices (b) of the Wess-Zumino model.}
\label{figWZ}
\end{figure}

So far we have seen an example where the MHV vertex expansion was trivial because there was only a single MHV amplitude ($\lambda \, \phi^4$ theory), and an example where there was an infinite tower of MHV amplitudes and the MHV vertex expansion was very powerful (\,(super) Yang-Mills theory). Going one step up in complexity from $\lambda \, \phi^4$, we add fermions and let them interact with the scalars via Yukawa-couplings. With supersymmetry, this gives the Wess-Zumino model with massless scalars and fermions. As we will see shortly, one interesting feature of this model is that it only has a finite number of MHV amplitudes. We consider the Wess-Zumino model with $N$ chiral superfields, $\Phi_a=z_a+\sqrt{2} \theta f_a+\theta^2 F_a$, a canonical K\"ahler potential, and the superpotential $\mathcal{W} = \tfrac{1}{6}g_{abc} \Phi_a \Phi_b \Phi_c$. The couplings $g_{abc}$ are fully symmetric and repeated indices are summed from 1 to $N$. The scalars $z_a$ have 4-point interactions $\tfrac{1}{4}g_{abx} g_{cdx}^* z_a z_b \bar{z}_c \bar{z}_d$, and they interact with the  fermions $f_a$
via Yukawa couplings $\tfrac{1}{2}g_{abc}\, z_a f_b f_c + \mathrm{h.c.}\,$. The table in figure~\ref{figWZ}(a) summarizes helicity, $U(1)_R$-charge, and supersymmetry $\alpha, \beta$ assignments (introduced in \reef{abstates}) of the states.

\vspace{2mm}
\noindent {\bf N$^k$MHV classification:~~}
As described in section~\ref{AllToMHV}, the external states of an N$^k$MHV amplitude are $(k+2)$ $\alpha$-states and $(n-k-2)$ $\beta$-states. If $n_x$ denotes the number of particles of type $x$ among the $n$ external states, we have (see table in figure \ref{figWZ}(a))
\bea
   \text{N$^k$MHV:} ~~~~~~
   \#\a\text{'s} \,=\,n_f + n_{\bar{z}} = k+2\, ,~~~~~~
   \#\b\text{'s} \,=\,n_{\bar{f}} + n_z = n-k-2\, .
\eea
Also, the sum of the $R$-charges must vanish: $- n_f -2 n_{\bar{z}} +n_{\bar{f}} + 2n_z = 0$. These three equations imply
\bea
   \label{NkWZ}
   \text{N$^k$MHV:}~~~~
   n_z = 6 - n_f + 3 k - n\, ,~~~~
   n_{\bar{z}} = k+2- n_f \, ,~~~~
   n_{\bar{f}} =  2n+ 4(k-2) + n_f  \, .
\eea
Note that since $n_z\ge 0$, there can be no N$^k$MHV amplitudes with more than $n_{\rm max}=6+3k$ external states. In particular, the {\it MHV sector} contains only amplitudes with $n=3,\dots,6$ external states. There are 7 such amplitudes: their MHV vertices are listed in figure  \ref{figWZ}(b) and explicit expressions are given in table \ref{WZmhv}.%
\footnote{The MHV amplitudes can be computed with BCFW recursion relations derived from a $[f,\bar{f}\>$ shift. We spare the reader the proof of the validity of this BCFW shift and the details of the computation of the MHV amplitudes from BCFW. }
Note that the MHV amplitudes are holomorphic in angle brackets.

\begin{table}[t]
\bea
  \< f_{a_1} f_{a_2}  z_{a_3} \> &=& - g_{{a_1}{a_2}{a_3}} ~\<12\> \, ,\\[1mm]
  \hline\phantom{{}^{\Bigl(}}
    \<  f_{a_1}  f_{a_2}  \bar{f}_{a_3}  \bar{f}_{a_4} \>
  &=&
  g_{{a_1}{a_2}x} \, g_{{a_3}{a_4}x}^* ~\frac{\<12\>}{\<34\>} \, ,\nonumber\\[1mm]
  \<  f_{a_1}    z_{a_2} \bar{f}_{a_3} \bar{z}_{a_4} \>
  &=&-
  g_{{a_1}{a_2}x} \, g_{{a_3}{a_4}x}^* ~\frac{\<14\>}{\<34\>} \, , \nonumber\\[1mm]
  \<    z_{a_1} z_{a_2} \bar{z}_{a_3}  \bar{z}_{a_4}\>
  &=&
  - g_{{a_1}{a_2}x} \, g_{{a_3}{a_4}x}^*  \, , \\[1mm]
  \hline\phantom{{}^{\Bigl(}}
   \nonumber
  \<  \bar{z}_{a_1}  \bar{z}_{a_2}  z_{a_3} \bar{f}_{a_4} \bar{f}_{a_5} \>
  &=&
   g_{{a_1}{a_2}x}^* \, g_{{a_3}xy}\, g_{{a_4}{a_5}y}^*  \frac{1}{\<45\>}
  - g_{{a_1}{a_4}x}^* \, g_{x{a_3}y}\, g_{{a_2}{a_5}y}^*  \frac{\<12\>}{\<14\>\<25\>} \\
  &&+ g_{{a_1}{a_5}x}^* \, g_{{a_3}xy}\, g_{{a_2}{a_4}y}^*  \frac{\<12\>}{\<15\>\<24\>}
  \, , \nonumber\\[1mm]
  \<  \bar{f}_{a_1}   \bar{z}_{a_2}f_{a_3}  \bar{f}_{a_4} \bar{f}_{a_5}  \>
  &=&
   -g_{{a_2}{a_1}x}^* \, g_{{a_3}xy}\, g_{{a_5}{a_4}y}^*
   \frac{\<23\>}{\<45\>\<12\>}
   + \mathcal{P}(3,4,5)\, , \\[1mm]
  \hline\phantom{{}^{\Bigl(}}
  \<  \bar{z}_{a_1}  \bar{z}_{a_2}  \bar{f}_{a_3} \bar{f}_{a_4} \bar{f}_{a_5} \bar{f}_{a_6} \>
  &=& g_{{a_1}{a_3}x}^* \, g_{{a_2}{a_4}y}^* \, g_{{a_5}{a_6}w}^* \, g_{xyw}
  ~\frac{\<12\>}{\<13\>\<24\>\<56\>} \nonumber\\
  &&- g_{{a_1}{a_2}x}^* \, g_{{a_3}{a_4}y}^* \, g_{{a_5}{a_6}w}^* \, g_{xyw}
  ~\frac{1}{\<34\>\<56\>}
  + \mathcal{P}(3,4,5,6)\,,\label{eq:zBzBfBfBfBfB}
 \eea
\caption{\small The seven MHV amplitudes of the Wess-Zumino model. $\mathcal{P}$ denotes a sum over \emph{inequivalent} permutations of the momentum and flavor labels of the listed particles, with a minus sign for interchanges of fermions. The amplitudes are not color-ordered.}
\label{WZmhv}
\end{table}

\vspace{2mm}
\noindent {\bf MHV vertex expansion:~~}
We use \reef{falloff} to find the large-$z$ behavior of the amplitudes under an anti-holomorphic shift. The couplings $g_{abc}$ are dimensionless, giving $c=0$, and the sum of helicities in an N$^k$MHV amplitude is $\sum_i h_i = \frac{1}{2}(n_{\bar{f}} -n_f) =  n+ 2(k-2)$, as can be seen from the table in figure \ref{figWZ}(a) and \reef{NkWZ}. Hence \reef{falloff} gives
\bea
   \hat{A}_n^\text{N$^k$MHV}(z) ~\sim~
 \frac{1}{z^k}
~~~~~\text{for}~~~~z \to \infty \, .
\eea
The four conditions of section~\ref{AllToMHV} are all satisfied, so the all-line shift recursion relations imply the validity of the MHV vertex expansion for all non-MHV amplitudes of the Wess-Zumino model. It is instructive to see it at work in the following.

\vspace{2mm}
\noindent {\bf Comparison of Feynman diagrams with the MHV vertex expansion:~~}
Pure scalar tree amplitudes behave exactly as in $\phi^4$-theory (section~\ref{sec:phiFourth}), so let us exchange a scalar
pair $\bar{z} z$ in a pure scalar amplitude with a fermion pair $\bar{f} f$. This does not change the N$^k$MHV level. The Feynman diagram expansion for the 6-point NMHV amplitude $\<z z \bar{z}\bar{z} f \bar{f}\>$ takes the schematic form
\begin{equation}
  \<z z \bar{z}\bar{z} f \bar{f}\>~=~
  \raisebox{-1pt}{
  \parbox[c]{4cm}{\includegraphics[width=3.8cm]{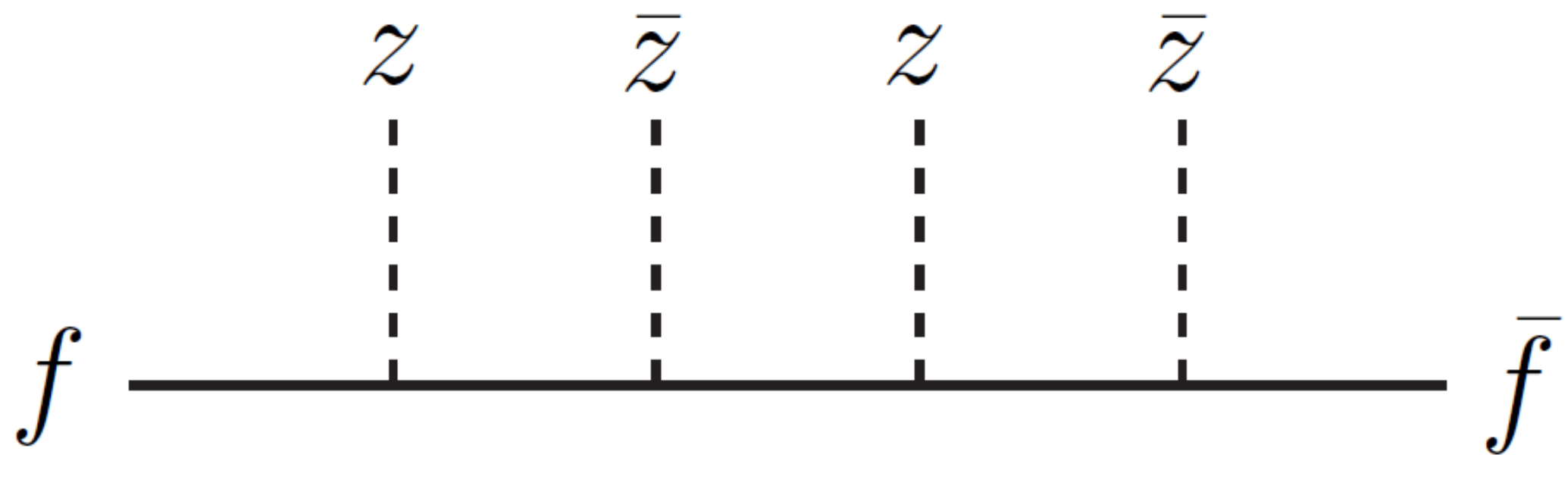}}}
  \!\!+
  \parbox[c]{3.2cm}{\includegraphics[width=3.1cm]{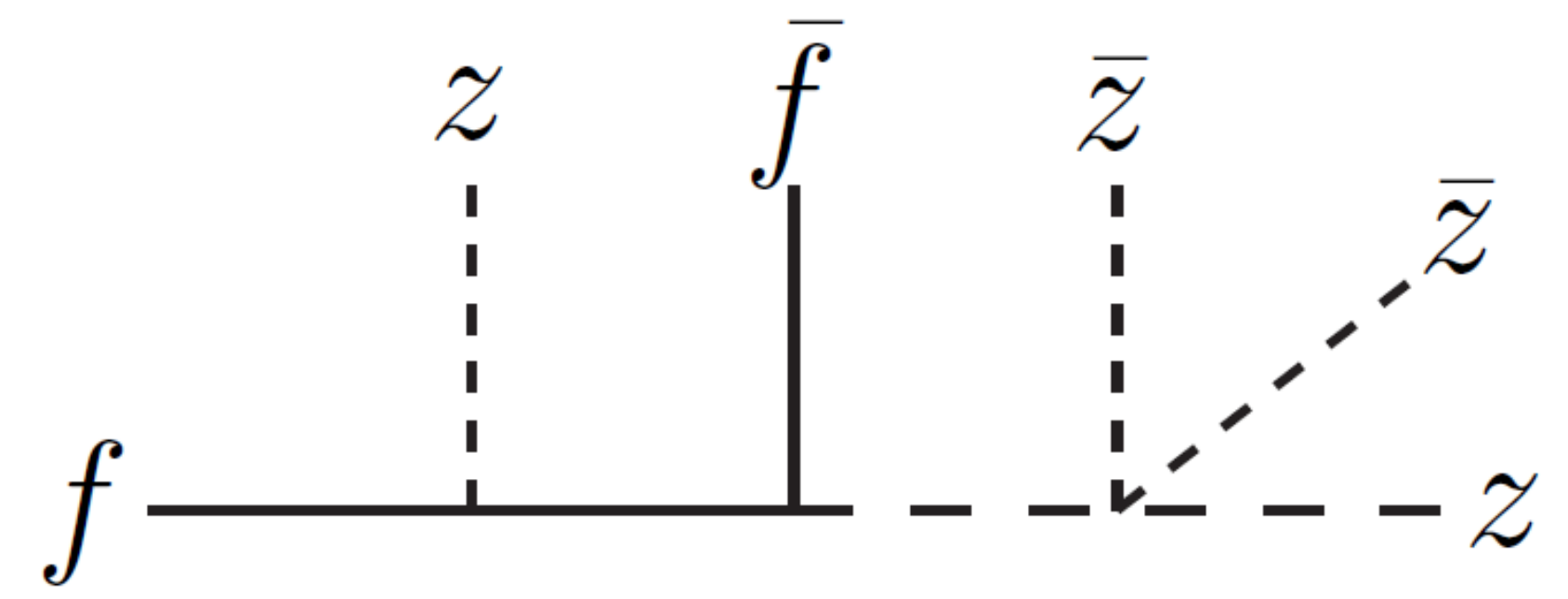}}
  +
  \parbox[c]{3.1cm}{\includegraphics[width=3.1cm]{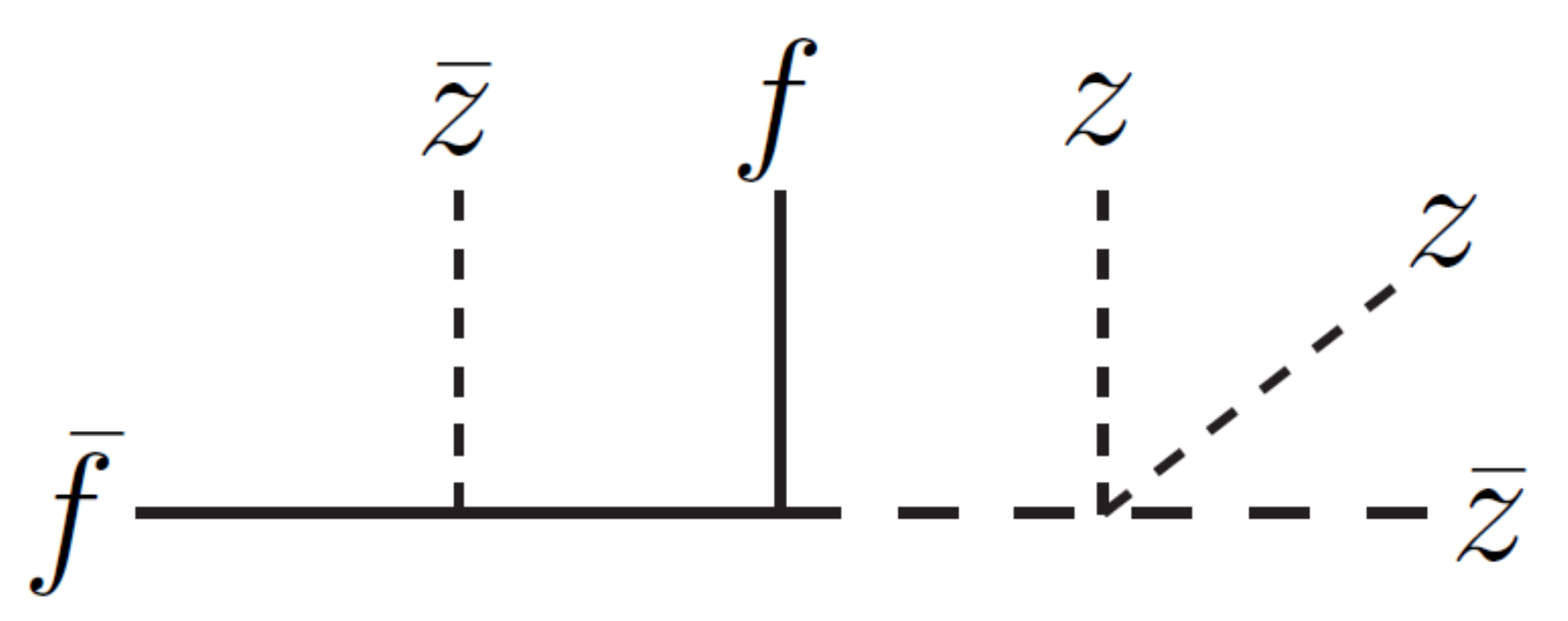}}\,.
\end{equation}
The MHV vertex expansion of the same amplitude is given by
\begin{equation}
\begin{split}
  \<z z \bar{z}\bar{z} f \bar{f}\>~=~
  \parbox[c]{2.8cm}{\includegraphics[width=2.7cm]{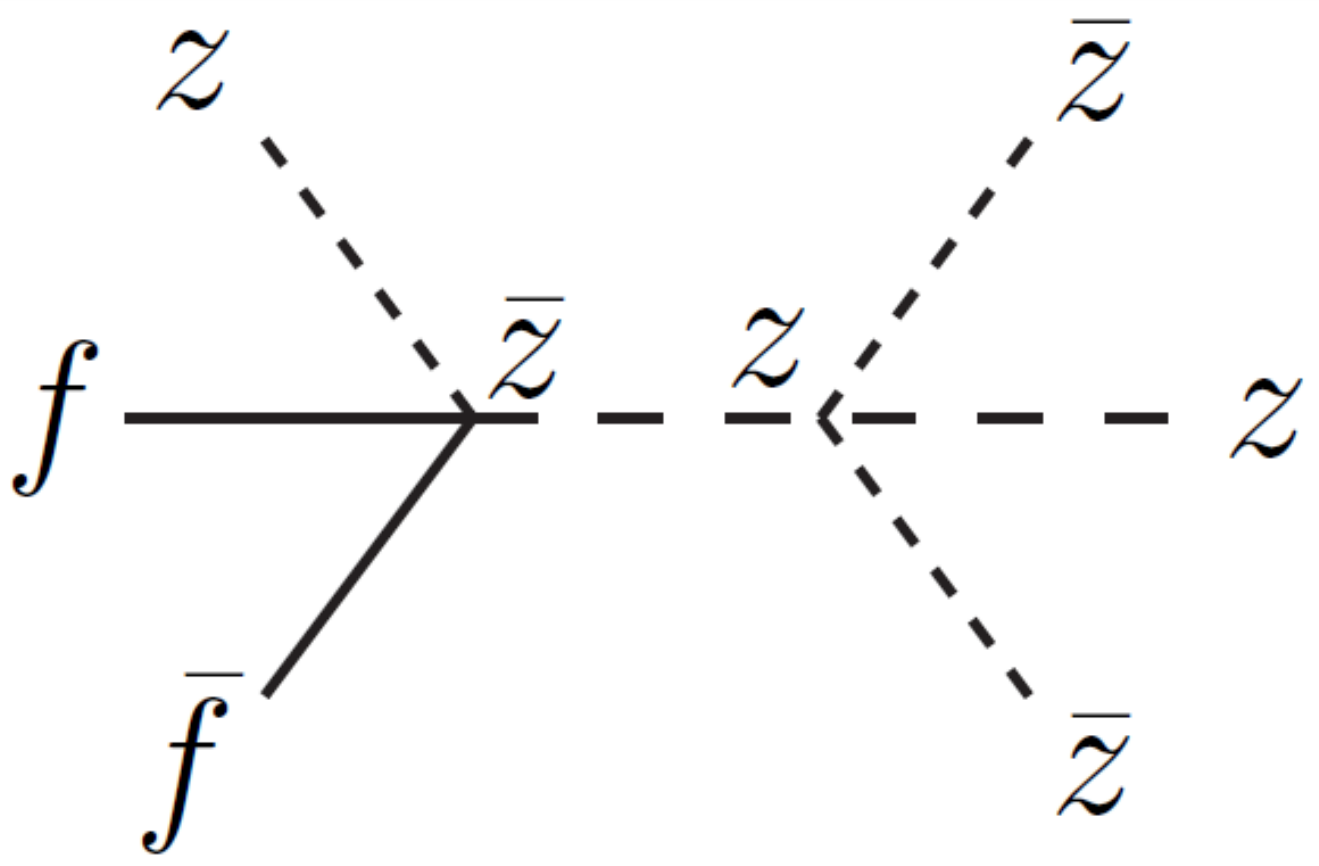}}
  \,+~
  \parbox[c]{2.8cm}{\includegraphics[width=2.7cm]{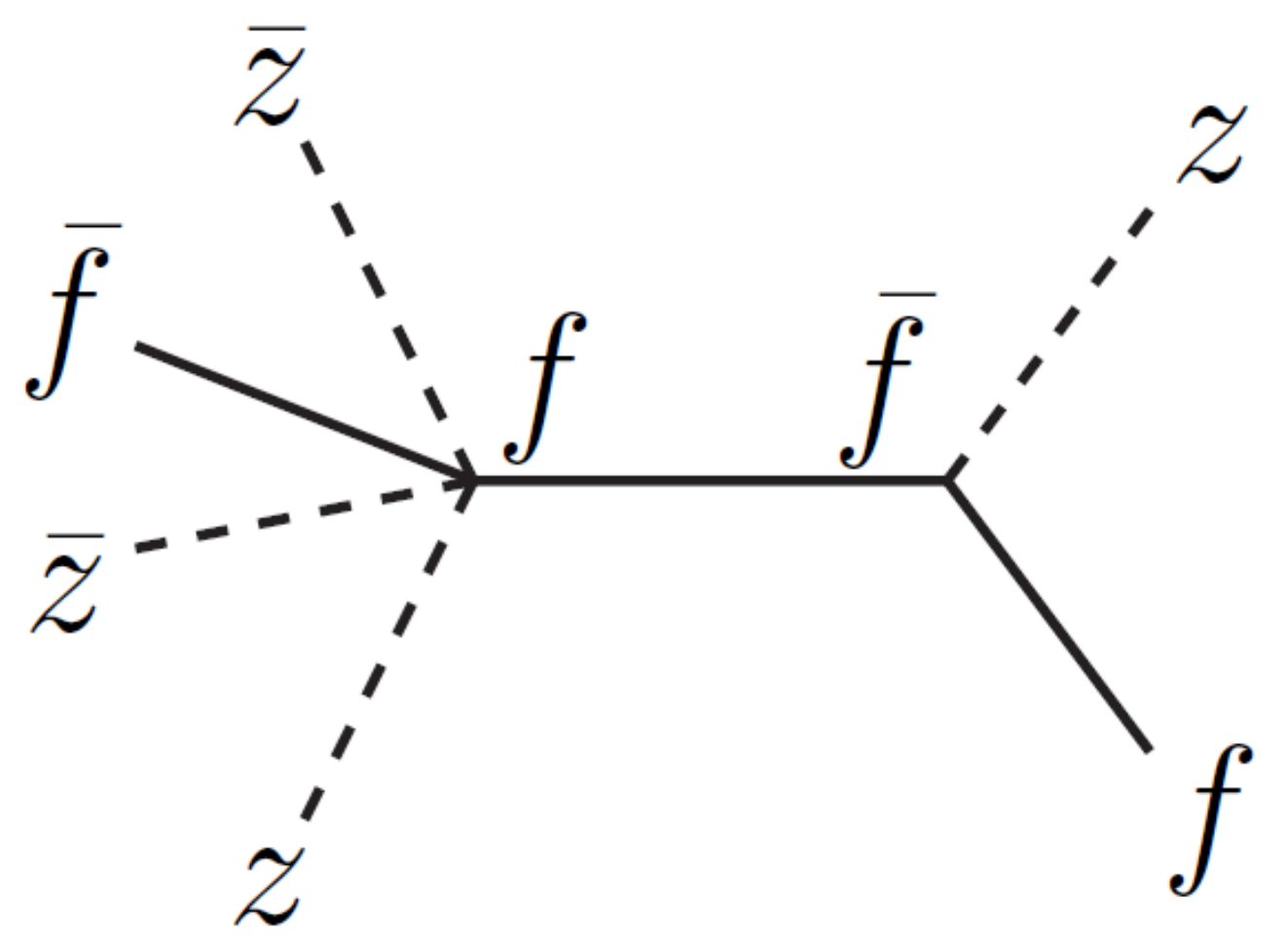}}
  \,+~
  \parbox[c]{2.8cm}{\includegraphics[width=2.7cm]{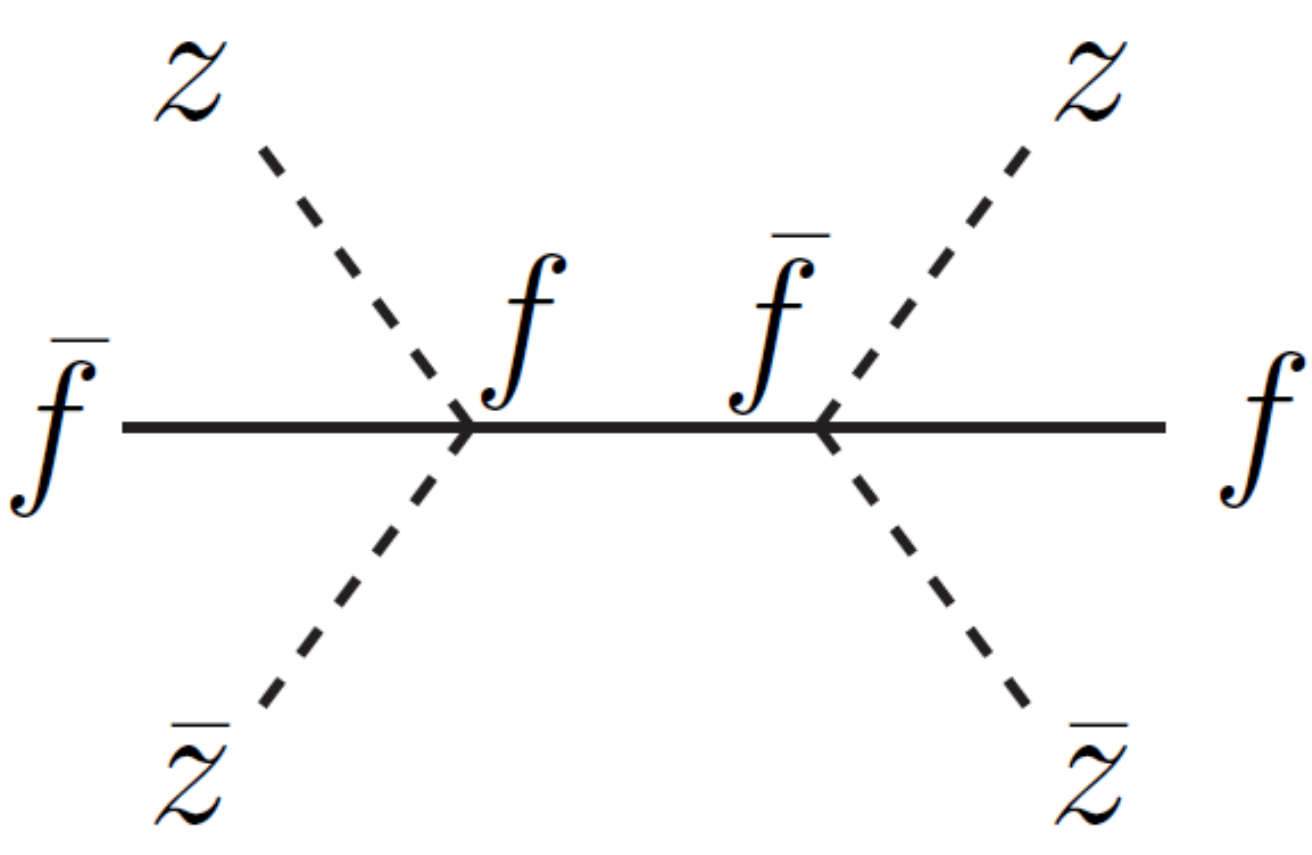}}
\end{split}
\end{equation}
plus diagrams obtained by replacing $z\leftrightarrow \bar{z}$ in the first diagram and $f\leftrightarrow \bar{f}$ in the second. There are more MHV vertex diagrams than Feynman diagrams. This has to do with the different ways Feynman diagrams can be reorganized into on-shell MHV blocks. Let us illustrate this explicitly in the simpler example of a 6-fermion amplitude.

\begin{figure}[t]
\centering
 \includegraphics[width=\textwidth]{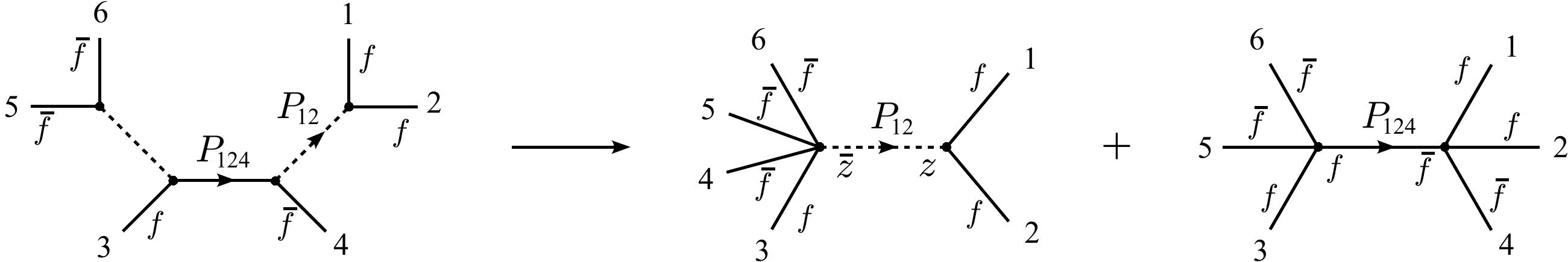}\vspace{.5cm}
\begin{displaymath}
\hspace{1cm} -\frac{\<3|P_{124}|4]}{P^2_{124}[12]\<56\>} \hspace{2.3cm}= \hspace{1.35cm} \frac{\<3|P_{12}|X]}{\<4|P_{12}|X][12]\<56\>}\hspace{.7cm}+\hspace{.7cm}\frac{-\<3|P_{124}|X]\<12\>}{\<4|P_{124}|X]\<56\>P^2_{124}}\,.
\end{displaymath}
\caption{\small A Feynman diagram (left) and the corresponding MHV vertex diagrams (right) of the  NMHV amplitude $\<fff\bar{f}\bar{f}\bar{f}\>$.}
\label{fig:WZfig6Fermion}
\end{figure}

Only one type of Feynman diagram contributes to the 6-fermion NMHV amplitude $\< f f f \bar{f} \bar{f} \bar{f} \>$, as displayed together with its value in figure~\ref{fig:WZfig6Fermion}. The full amplitude is the sum of diagrams obtained from cyclic permutations of lines $(1,2,3)$ and $(4,5,6)$. There are two ways to `chop' the propagators in the Feynman diagram in figure \ref{fig:WZfig6Fermion} to get MHV vertices:%
\footnote{Cutting $P_{56}$ gives an anti-MHV 3-vertex.}
cut $P_{12}$ or $P_{124}$. The results, given in figure \ref{fig:WZfig6Fermion}, are exactly the two types of diagrams that appear in the MHV vertex expansion of $\< f f f \bar{f} \bar{f} \bar{f} \>$. The sum of these two diagrams is independent of $|X]$ and agrees with that of the Feynman diagram.

The MHV vertex expansion gives an alternative on-shell formulation of the tree-level Wess-Zumino model. We expect that there exists a corresponding MHV vertex Lagrangian \cite{Mansfield:2005yd, Ettle:2006bw, Feng:2006yy} in which the anti-MHV 3-vertex $\bar{z} \bar{f} \bar{f}$ is absent at the cost of
having
7 fundamental MHV interactions.

\subsection{Gluons $\to$ Higgs fusion}
\label{sec:GluonHiggsFusion}
In the Standard Model, the Higgs $h$ interacts with gluons (or photons) through 1-loop diagrams with a fermion running in the loop. Gluon fusion processes $gg \to h$ are expected to be the dominant source of Higgs-production at the LHC. The leading contribution comes from the top quark loop. If the Higgs mass is below the threshold for $t\bar{t}$ creation, $m_h < 2 m_t$, integrating out the heavy quark gives an effective description of gluon fusion in terms of the dimension-5 operator
\bea
  \label{hFF}
  \frac{\hg}{2} \, h \, {\rm Tr}\,F_{\mu\nu} F^{\mu\nu}\, .
\eea
The coupling constant is $\hg= \alpha_s/(6 \pi v)$, with $v \approx 246$ GeV. Amplitudes with $m$ insertions of the operator \reef{hFF} are suppressed as $(\sqrt{s}/v)^m$, so we restrict our attention to tree amplitudes with a single insertion of \reef{hFF}. The MHV vertex expansion has been used in the literature to calculate these amplitudes \cite{Dixon:2004za, Badger:2004ty, Berger:2006sh, Badger:2007si, Dixon:2009uk, Badger:2009hw}.   Here we apply our general results of sections \ref{AllLineShifts} and \ref{AllToMHV} to justify the validity of the MHV vertex expansion.

The first step is to embed the operator \reef{hFF} into an ${\cal N}=1$ supersymmetric theory \cite{Dixon:2004za, Berger:2006sh} so we can define an MHV classification. To this end we introduce a vector supermultiplet and a chiral superfield $\Phi$ whose lowest component  is a complex scalar $z$. The real part of $z$ is the Higgs, $h~=~\tfrac{1}{2}(z+\bar{z})$. The effective operator $\Phi\,{\rm Tr}\, W_\a W^\a$ + h.c. yields bosonic component operators ${\rm Re} \, z\, {\rm Tr} \, F^2 = h \, {\rm Tr}\, F^2$ and ${\rm Im} \, z\, {\rm Tr}\, F \tilde{F}$. It is useful to express the amplitudes in terms of the complex scalar $z$ and its conjugate $\bar{z}$, since they couple holomorphically/anti-holomorphically to the negative/positive helicity gluons \cite{Dixon:2004za, Berger:2006sh}:
\begin{equation}
\begin{split}
  &~\< -- z \> \,=\, -  \hg\, \<12\>^2\, ,~~~~~\< ++ \bar{z} \> \,=\, -\hg\, [12]^2\,,\\[1mm]
  &\< ++ z \> \,=\,  \< --\bar{z} \> \,=\,\< +- z \> \,=  \< -+\bar{z} \> \,=\, 0\,.
\end{split}
\end{equation}
Higgs amplitudes are linear combinations of amplitudes involving $z$ and $\bar{z}$, for example\\
$\<--h\> = \<--z\>+\<--\bar{z}\> = \<--z\>$.

Under the action of the SUSY charges \reef{abstates}, gluons $(A^\pm)$ and scalars ($z$, $\bar{z}$) are classified as
\begin{equation}
    \alpha\text{-states: }~\bar{z}\,,~ A^-\,
    ,
    \qquad\beta\text{-states:~}~ z\,,~ A^+\,.
\end{equation}
When all particles are massless, supersymmetry guarantees that the ``ultra helicity violating" (UHV) amplitudes $\<\a \dots \a\>$ and $\<\a \dots \a \b\>$ vanish. However, when massive particles are involved, amplitudes $\<\a \dots \a \b\>$ are generically non-vanishing. The gluons are massless, but the Higgs is massive. With one external scalar, the UHV amplitudes \cite{Dixon:2004za, Berger:2006sh} are
\begin{equation}
\label{UHVhgf}
\< \pm + \dots + z \> ~=~0  \, , \qquad~~
\< + + \dots +  h \>~=~\< + + \dots +  \bar{z} \> ~=~ \hg\,  \frac{m_h^4}{\mathrm{cyc}(1,n\dash1)}  \, ,
\end{equation}
with $\mathrm{cyc}(1,n\dash1)$ defined as the cyclic product of the angle brackets involving only the gluon momenta. The vanishing of $\< + + \dots + z \> = \< \b  \b \dots \b \>$ follows from the SUSY Ward identities, while the vanishing of $\< - + \dots + z \>$ can be proven inductively using BCFW. The compact formula for the UHV amplitude $\< + + \dots +  h \>$ was found in \cite{Dixon:2004za, Berger:2006sh}; it can also be derived using a holomorphic all-line shift, but we will not include the details here.

Let us now move on to the MHV sector.  Recalling that the external states of MHV amplitudes are 2 $\a$-states and $(n-2)$ $\b$-states, we note that (up to permutations of gluons) the bosonic MHV amplitudes are $\<--+\ldots+z\ldots z\>$, $\<-+\ldots+\bar{z}\, z\ldots z\>$, and $\<+\ldots+\bar{z}\,\bar{z}\,z\ldots z\>$. The gluons are color-ordered, while the positions of scalars are arbitrary. With only one insertion of \reef{hFF}, the only non-vanishing MHV amplitudes are
\be
 \text{MHV:}~~~~~ \<--+\ldots+z\> ~~~~ \text{and}~~~~
 \<-+\ldots+\bar{z}\>\,.
\ee

The MHV amplitude $\<--+\ldots+z\>$ was calculated in~\cite{Dixon:2004za, Badger:2004ty} and found to take the same form as the Parke-Taylor amplitude,
\begin{equation}\label{MHVhgf}
  \< + \dots -_i \dots -_j  \dots + z \> ~=~ -\hg \, \frac{\<ij\>^4}{\mathrm{cyc}(1,n\dash1)}  \,.
\end{equation}
This formula is valid both for massive and massless scalars, and it can be proven recursively using BCFW. Together with the usual gluon MHV amplitudes, the single-scalar amplitudes \reef{MHVhgf} give an MHV vertex expansion for N$^k$MHV amplitudes of the form
\begin{equation}\label{NkMHVz}
  \text{N$^k$MHV:}~~~~~~
  \< - \ldots-+ \ldots +\,z\>\,.
\end{equation}
To justify the MHV vertex expansion, we first use \reef{falloff} to show that the anti-holomorphic all-line shift recursion relations are valid.%
\footnote{For a massive Higgs boson, the justification for using \reef{falloff} is given in section~\ref{sec:MassiveParticles}.}
With one insertion of \reef{hFF} we have $c=-1$, and $k\!+\!2$ negative helicity gluons and $n\!-\!k\!-\!3$ positive helicity gluons give $\sum_i h_i = n -2k -5$. By \reef{falloff}, the amplitudes \reef{NkMHVz} therefore fall off as $1/z^k$ for large $z$. Secondly, even if all four conditions in section \ref{AllToMHV} are satisfied, there is a potential obstacle since the massive scalar could result in UHV amplitudes appearing in the all-line recursion relations. However, the only UHV amplitudes that could appear in the expansion of~(\ref{NkMHVz}) are $\< \pm + \dots + z \>$, but they vanish according to~(\ref{UHVhgf}). Therefore, the only possible subamplitudes in the all-line shift recursion relations are the holomorphic MHV amplitudes (\ref{MHVhgf}), pure-gluon amplitudes, and lower-point amplitudes of the form~(\ref{NkMHVz}). Thus, despite the massive Higgs boson in this theory, the MHV vertex expansion is rigorously justified.

As a final comment, let us note that there is a simple formula for the MHV amplitudes with $\bar{z}$ when the scalars are massless:
\begin{equation}\label{mpbarz}
  \< - + \ldots +\, \bar{z} \> ~=~ \frac{\hg}{\mathrm{cyc}(1,n\dash1)}
  \sum_{j=2}^{n-1}
  \frac{\<1|p_j.p_{\bar{z}}|1\>^2}{s_{{\bar{z}} j}}
    \, ,  \hspace{1cm}\text{(for $m_h=0$\,)}\, .
\end{equation}
This can be proven inductively, using a BCFW shift $[-,+\>$ of adjacent gluon lines.

\subsection{$F^m$ and $R^m$ operators}

How is on-shell constructibility affected when a higher-derivative interaction $ \alpha_m D^{2q}F^m$ is added to the Yang-Mills Lagrangian? Or when $\b_m D^{2q}R^m$ is added to gravity? Such operators appear in the open and closed string effective action,%
\footnote{For recent work on recursion relations for string amplitudes, see \cite{Boels:2008fc,Cheung:2010vn,Boels:2010bv}.}
and they can also be considered candidate counterterms for UV divergences in loop-amplitudes. In general, these higher-dimensional operators may or may not be supersymmetrizable. In this section we consider matrix elements with a single insertion of operator $F^m$ to Yang-Mills theory or of $R^m$ to Einstein gravity.%
\footnote{See \cite{Brodel:2009hu,Elvang:2010jv,Elvang:2010kc,Beisert:2010jx} for recent analyses of the matrix elements of supersymmetrizable gravity operators.}
Information about the particular index contractions or trace structure is not needed for our analysis. On-shell constructibility for the more general versions of these operators is discussed in section \ref{sec:DiscussionOfConstructibility}.

Matrix elements with a single insertion of $\alpha_m F^m$ are denoted by $\< \dots \>_{F^m}$. These are proportional to the coupling $\alpha_m$ which has mass dimension ${4-2m}$ (see table \ref{tab:cValues}), and hence $c=4-2m$. Suppose the operator $F^m$ is not supersymmetrizable. Then amplitudes with less than two positive-helicity gluons do not have to vanish; let us in particular focus on the ultra helicity violating (UHV) all-minus amplitudes. The general formula \reef{falloff} shows that under an anti-holomorphic shift, $\<\hat{-}\hat{-}\dots \hat{-}\>_{F^m} \sim z^{m-n}$ for large $z$. Hence for $n>m$, the all-line shift recursion relations allow us to construct the $n$-point amplitudes $\<- -\dots -\>_{F^m}$. This is the strongest constructibility that one can expect: the leading interaction of $F^m$ is $m$-point, so the amplitudes with $n<m$ do not have insertions of $F^m$. For $n=m$ we need the input from the $m$-point vertex since its information cannot possibly be constructed from the lower-point Yang-Mills interactions. When $n>m$ there are local interaction terms in the non-linear completion of $F^m$, but these are inferred from gauge invariance and do not contain independent information. Hence, it makes sense that the all-minus $n$-point amplitudes with $n>m$ can be computed recursively.

As an explicit example, consider the operator
\begin{equation}
    \alpha_3\,
    F^3
= \alpha_3\, \mathrm{tr}\, F_\mu{}^\nu \, F_\nu{}^\lambda \, F_\lambda{}^\mu\,.
\end{equation}
Its leading 3-point interaction gives%
\footnote{Note that $\< - - + \>_{F^3}=0$.}
$\< - - - \>_{F^3} = \alpha_3 \<12\> \<23\> \<31\>$. This is the unique spinor product of mass dimension 3 with the correct little-group scaling. Together with the usual on-shell Yang-Mills amplitudes, $\< - - - \>_{F^3}$ is  the  only input needed to construct all the higher-point matrix elements $\< - - \dots - \>_{F^3}$. For example, the all-line shift recursion diagrams of $\< - - - \,- \>_{F^3}$ consist of a $\< - - - \>_{F^3}$ subamplitude, together with a standard Yang-Mills theory MHV subamplitude $\< - - + \>$. Summing over cyclic permutations, we find
\bea
  \< - - - \,- \>_{F^3}
  ~=~ \alpha_3\sum_{\mathcal{P}_c(1234)} \<12\> \<2 P_{12}\> \<P_{12}1\>    \frac{1}{P_{12}^2} \frac{\<34\>^3}{\<3 P_{12}\>\<P_{12} 4\>}
  ~=~ 2\alpha_3 \frac{s\,t\,u}{[12][23][34][41]}\, .
\eea
In the last step we carried out the cyclic sum to determine the $|X]$-independent result. The final expression is obviously cyclically invariant. The example shows that the operator $F^3$ generates ultra helicity violating amplitudes. These are not permitted in a supersymmetric theory, and we conclude that $F^3$ is not supersymmetrizable; a well-known result.

Next, consider the gravity operator $\beta_m\, R^m$ (with $m\ge 3$) constructed by contracting $m$ Riemann tensors. Matrix elements $\< \dots \>_{R^m} $  with a single insertion of $R^m$ have $c=4-n-2m$ (see table \ref{tab:cValues}), and hence $\<\hat{-}\hat{-}\dots \hat{-}\>_{R^m} \sim z^{m-n}$ for large $z$. The all-line shift recursion relations are therefore valid for $n>m$, as in the case of $F^m$. Let us illustrate the recursion relations for $R^3$. The 3-point matrix element is unique,
\bea
 \< - - - \>_{R^3} ~=~ \beta_3\<12\>^2 \<23\>^2 \<31\>^2
 ~\propto~
 \big(\< - - - \>_{F^3}\big)^2\, .
\eea
The all-line shift recursion relations give
\bea
  \< - - - \,- \>_{R^3}
  &=&\nonumber
  \tfrac{1}{4}\beta_3
  \sum_{\mathcal{P}(1234)}
  {\<12\>^2\<2P_{12}\>^2\<P_{12}1\>^2
  \frac{1}{P_{12}^2}
  \frac{\<34\>^6}{\<4P_{12}\>^2\<P_{12}3\>^2} }\\
  &=&
  \tfrac{1}{4}\beta_3
  \sum_{\mathcal{P}(1234)}
  \frac{\<12\>^5 \<34\>^2 [1X]^2 [2X]^2}{[12][3X]^2 [4X]^2} \, ,~~
\eea
where the sum is over all permutations of momenta 1,2,3,4 and the factor of $1/4$ corrects for overcounting. We have verified $|X]$-independence numerically; in fact, a KLT-like relation holds:
\bea
  \<1^-,2^-,3^-,4^-\>_{R^3}
  \propto \, s_{12}\,  \<1^-,2^-,3^-,4^-\>_{F^3} \<1^-,2^-,4^-,3^-\>_{F^3} \, .
\eea
Originally, the matrix element $\< - - - \,- \>_{R^3}$ was calculated by van Nieuwenhuizen and Wu \cite{vanNieuwenhuizen:1976vb} using a much more involved Feynman diagram calculation. The non-vanishing of  $\< - - - \,- \>_{R^3}$
 shows
that $R^3$ is not supersymmetrizable \cite{Grisaru:1976nn}.

\subsection{Ultra  helicity violating 1-loop amplitudes}
We have developed all-line shift recursion relations for tree amplitudes, but they can also be applied to the rational part of loop amplitudes. Here, as an example, we verify inductively the formula \cite{Bern:1993qk,Mahlon:1993si}
\bea\label{UHV1loop}
  &&
  R_n^\text{1-loop} (- -\dots -) ~=~ Z\,\frac{f_n(1,n)}{\overline{\text{cyc}}[1,n]}\, ,\nonumber
  \\[1mm]
  \label{g_n}
  &&f_n(1,n) ~\equiv \!\!\!\!\!\!\!\sum_{1\le i_1 < i_2< i_3 < i_4 \le n}\!\!\!\!\!\!
    [i_1 i_2] \<i_2 i_3\> [i_3 i_4] \<i_4 i_1\> \, ,\qquad
    \overline{\text{cyc}}[1,n]~\equiv~[12][23]\cdots[n1]
\eea
for the planar contribution to the (color-ordered) ultra helicity violating (UHV) 1-loop amplitude in QCD. $Z$ is a constant containing the coupling and group-theory factors, and it is easy to see that $f_n$ is cyclically invariant. Recursive derivations of~(\ref{UHV1loop}) based on BCFW were given in \cite{Bern:2005hs,Bern:2005ji}.%
\footnote{Risager's anti-holomorphic 3-line shifts \cite{Risager:2005vk} were used for UHV 1-loop gravity amplitudes in \cite{Brandhuber:2007up}.}
We show in this section that the all-line shift provides a very simple new recursive proof of this result.

\noindent {\bf Validity of all-line shift:~~}
Under an anti-holomorphic all-line shift, the large-$z$ falloff of amplitudes with $n$ negative helicity gluons and dimensionless couplings is given by \reef{falloff} as $\hat{R}_n^\text{1-loop} (\hat{-} \hat{-}\dots \hat{-}) \sim z^{2-n}$. Hence, assuming that $R_n^\text{1-loop}(- - \dots -)$ is a rational function, the all-line shift recursion relations are justified for any $n\ge4$.

\noindent {\bf Inductive derivation:~~}
For $n=5$, the result was first established in~\cite{Bern:1993mq}. We proceed inductively and assume that~(\ref{UHV1loop}) is valid for $n-1$ external gluons. The diagrams in an all-line shift recursion relation for $\hat{R}_n^\text{1-loop} (- -\dots -)$ must consist of a 1-loop and a tree-level subamplitude. Since all external states are negative helicity gluons, the tree-level subamplitude is only non-vanishing when it has a total of three lines; up to cyclic permutations, it must therefore be of the form $\hat{A}^\text{tree}_3 (({n\mhat 1})^-\!\!, \hat n^- \!\!, -\hat{P}^+ )$, where  $\hat{P}=\hat{P}_{n\dash1,n}$. This leaves $\hat{R}^\text{1-loop}_{n-1} (\hat{1}^-\!\!, \dots (n\mhat2)^-\!\!,\hat{P}^-)$ as the other subamplitude. The resulting diagram $D$ is then given by
\bea\label{D}
 D ~=~
 ~\parbox[c]{4.25cm}{\includegraphics[width=4.25cm]{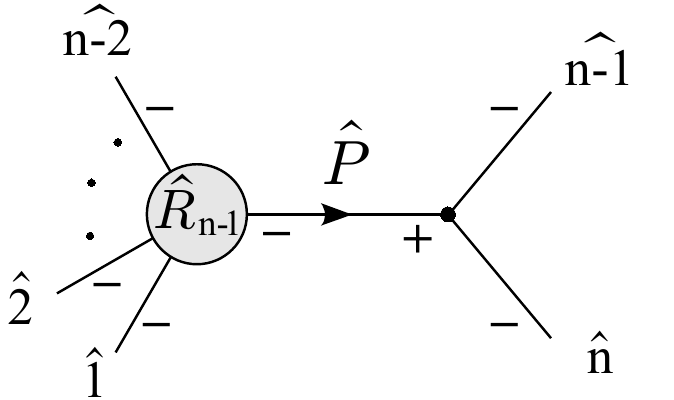}}~~=~~
  Z\,\,\frac{\hat{f}_{n-1}(\hat{1},\hat{P})}
  {\overline{\text{cyc}}[\hat{1},\hat{P}]}\times
  \frac{1}{P_{n\dash1,n}^2}\times \frac{\< n\!-\!1, n\>^3 }{\<n\hat{P}\>\<\hat{P}, n-1\> }\,,
\eea
where the shifted momenta are evaluated at the value of $z$ that puts the internal momentum $\hat P\equiv\hat{P}_{n\dash1,n}$ on-shell. From $|\hat{P}] \<\hat{P}|=|n\mhat 1] \<n\!-\!1|+|\hat n] \<n|$ and $[n\mhat1,\hat{n}] = 0$ it follows that
\bea
  \hat{f}_{n-1}(\hat{1},\hat{P})=\hat{f}_{n}(\hat{1},\hat{n})\,.
\eea
In the denominator of~(\ref{D}), the spinor products involving the internal line $\hat P$ give
\begin{equation}
      [n\mhat2,\hat{P}]\,[\hat{P}\,\hat 1]~\<n\hat{P}\>\,\<\hat{P},n-1 \>
 = \<n\!-\!1, n \>^2 \, [n\mhat2\, n\mhat1] \,[\hat n\,\hat1]  \, .
\end{equation}
It follows that
\begin{equation}
    D~=~Z\,\frac{\hat f_n(\hat 1,\hat n)}{[\hat 1\hat2]\cdots[n\mhat2,n\mhat 1]\,[n\!-\! 1,n]\,[\hat n\hat 1]}\,.
\end{equation}
We recognize $D$ as (minus) the all-line-shift residue at $\hat P^2_{n\dash1,n}=0$ of
\begin{equation}\label{Ragain}
    Z\,\frac{f_n(1,n)}{\overline{\text{cyc}}[1,n]}\,.
\end{equation}
Summing over all cyclically related diagrams then gives us the  residues of~(\ref{Ragain}) at $\hat P^2_{i,i\!+\!1}=0$ for any $i$. Since $\hat{R}_n^\text{1-loop} (- -\dots -)$ is constructible from an all-line shift it is completely determined by its residues, and it must therefore coincide with~(\ref{Ragain}). This completes the derivation of~(\ref{UHV1loop}).

\setcounter{equation}{0}
\section{Massive particles}\label{sec:MassiveParticles}
We extend the all-line shift recursion relations to amplitudes with massive particles. We introduce an ``anti-holomorphic'' all-line shift for massive particles and use a massive spinor-helicity formalism to determine the large-$z$ behavior; the
necessary machinery is presented in this section while further details are relegated to appendix \ref{sec:SpinorHelicityFormalism}. We illustrate massive all-line shift recursion relations with several examples.

\subsection{Massive all-line shifts}
\label{sec:mshift}
The constraints on consistent all-line shifts are strong, and it turns out that there is a unique way to define them. Begin with a shift of all the external states $i = 1,\dots,n$:
\begin{equation}\label{eq:genericalMomShift}
    p_i\to p_i+z \,\shk_i\,.
\end{equation}
We require that
\begin{itemize}
  \item External momenta stay on-shell: $\shk_i ^2=0$ and
  $\shk_i\cdot p_i=0$   for each $i=1,\dots,n$.
  \item Momentum is conserved: $\sum_{i=1}^n \shk_i=0$.
  \item
  All multi-particle invariants shift linearly in $z$: need $\shk_i \cdot  \shk_j=0$ to
  eliminate $\mathcal{O}(z^2)$-terms.
\end{itemize}
Up to conjugation, the unique way to satisfy these conditions is
\begin{equation}
  \label{massiveshift}
    p_i^{\da\b}~\to~\hat p_i^{\,\da\b}
    = p_i^{\da\b}+z\, d_i \, p_i^{\da\g}|X]_\g[X|^\b\,,
        \qquad \sum_{i=1}^n d_i\, p_i|X] =0\,.
\end{equation}
For a generic choice of the external momenta and reference spinor $|X]$, the constraint in~(\ref{massiveshift}) on the constants $d_i$ can be satisfied for $n\geq 4$ external lines.%
\footnote{We choose a sufficiently generic solution for the $d_i$ to avoid that the $\mathcal{O}(z)$-terms in multi-particle invariants cancel.}

If $p_i$ is massless, then $d_i \, p_i|X][X| = d_i \, |i\>[i X] \, [X|$; we recognize this as the conventional anti-holomorphic all-line shift \reef{allline} with $w_i = d_i [i X]$.

\subsection{Spinor-helicity formalism for massive particles}
\label{s:massSH}
There are two main approaches to spinor-helicity formalisms for massive particles (see, for example, Dittmaier~\cite{Dittmaier:1998nn}): one is based on identifying eigenvectors of the momentum matrix $p_{\alpha\dot{\beta}} = p_\mu \sigma^\mu_{\alpha\dot{\beta}}$ and gives Dirac spinor solutions that are eigenstates of the helicity operator. The other approach decomposes the time-like momenta $p_i$ along two light-like directions by introducing a null reference vector $q_i$ for each state. Our generalization of all-line shifts to amplitudes with massive external states is most naturally studied in the latter formalism.%
\footnote{See~\cite{Boels:2008du,Boels:2009bv,Boels:2010mj} for earlier applications of BCFW recursion relations with massive particles.}

Consider a massive momentum $p$ with $p^2=-m^2$. Following~\cite{Dittmaier:1998nn} we introduce a light-like reference vector $q$ and decompose $p$ as
\begin{equation}
    p~=~p^\perp-\frac{m^2}{2q\cdot p}\, q ~=~ p^\perp+\frac{m^2}{\<q|p|q]}\, q\,.
\end{equation}
The projection $p^{\perp}$ is null and has associated angle and square spinors $|p^\perp]$ and $\<p^\perp|$ defined as
\begin{equation}\label{pperps}
p^\perp=|p^\perp\>[p^\perp|\,, \qquad\text{with}\quad
    |p^\perp]=\frac{p |q\>}{\,\sqrt{\<q|p|q]}\,}\,, \qquad
    |p^\perp\>=\frac{ p|q]}{\,\sqrt{\<q|p|q]}\,}\,.
\end{equation}
Note that $|p^\perp\>^*=|p^\perp]$ for real momenta and real reference spinor $q$.

In the massless limit, the angle and square spinors differ from the usual conventions by $t |p\>$ and $t^{-1}|p]$ with $t= \sqrt{[qp]/\<pq\>}$, which can be compensated by a little-group scaling. Hence, to take the massless limit of a state in an amplitude, we can simply replace $|p^\perp\>$ and $|p^\perp]$ by $|p\>$ and $|p]$ as $m\to 0$.

\paragraph{Dirac fermions:}
The independent solutions to the massive Dirac equation \reef{Dirac}, fermions $\overline{u}_s$ and $u_s$ and anti-fermions $\overline{v}_s$ and $v_s$, are labeled by $s=\pm$. We choose these solutions to be eigenvectors of the helicity operator $\tilde{\Sigma}^\pm_{p;q}$ in the frame defined by $q$ and $p$. The explicit expression for $\tilde{\Sigma}^\pm_{p;q}$ is given in \reef{SigmaPM}. $\tilde{\Sigma}^\pm_{p;q}$ determines the ``{\bf {\emph q}-helicity}" $\tilde{h}_i = \pm \tfrac{1}{2}$ of the states. For example, the outgoing $\tilde{h}_i = \pm \tfrac{1}{2}$ anti-fermions are
\begin{equation}\label{doublebracketdef}
    |p\dsk \equiv v_+
    =\begin{pmatrix} |p^\perp]\\[1mm] \frac{im}{\sqrt{\<q|p|q]}}|q\> \end{pmatrix}\,, \qquad |p\dak \equiv  v_-=
    \begin{pmatrix} \frac{im}{\sqrt{\<q|p|q]}}|q]\\[2mm] |p^\perp\> \end{pmatrix}\, ,
\end{equation}
and they satisfy $\tilde{\Sigma}^+_{p;q}  |p\dsk = |p\dsk$ and $\tilde{\Sigma}^-_{p;q}  |p\dak = |p\dak$, while $\tilde{\Sigma}^-_{p;q}  |p\dsk = \tilde{\Sigma}^+_{p;q}  |p\dak =0$.

The solutions for the outgoing fermions $\dsb p| = -i \, \overline{u}_+$ and $\dab p | = i\, \overline{u}_-$ are given in \reef{outferm}. They have  {\emph q}-helicity $\tilde{h}_i = \pm \tfrac{1}{2}$, respectively. We distinguish incoming fermions from outgoing ones by a ``bullet'' on their bra-kets; for example, an incoming negative {\em q}-helicity fermion is denoted by $|p\dskB$. Crossing symmetry relates it to an outgoing positive {\em q}-helicity anti-fermion: $|(- p)\dskB  = |p\dsk$. Similarly, $|(- p)\dakB  = - |p\dak$. We can summarize the Feynman rules as
\be
\hspace{-5.5mm}
\label{Dfeyn}
\text{
\begin{tabular}{cccccccccc}
 &\underline{\sl Outgoing} & fermion & anti-fermion
 &&&
  \underline{\sl Incoming} & fermion & anti-fermion\\[2mm]
  &$\tilde{h}=-\frac{1}{2}$ & $\overline{u}_-  \lra\, \dab p|$ & $v_-\lra \, |p\dak$
  &&&
  $\tilde{h}=+\frac{1}{2}$ & $u_+ \lra\, -|p\dakB$ & $\overline{v}_+ \lra\, -\dabB p|$ \\[3mm]
  &$\tilde{h}=+\frac{1}{2}$ & $\overline{u}_+ \lra\, \dsb p|$ & $v_+ \lra\,|p\dsk$
  &&&
  $\tilde{h}=-\frac{1}{2}$ & $u_- \lra\,|p\dskB$ & $\overline{v}_- \lra\,\dsbB p|$\,.
\end{tabular}
}
\ee
Note that outgoing angle/square-spinor states have negative/positive  {\em q}-helicity while incoming angle/square-spinor states have the opposite, namely positive/negative, {\em q}-helicity. This is consistent with the crossing rules.

The familiar completeness relation $v_+ \overline{v}_+ + v_- \overline{v}_- = i \slashed{p} -m$, and the similar version with $u$'s, take the following form in bra-ket notation:
\bea
  \label{complete1}
  |p \dak \dsbB p|
  - |p \dsk \dabB p|  = \slashed{p} + im \, ,~~~~~~~~~~
  |p \dakB \dsb p|
  - |p \dskB \dab p|  = \slashed{p} - im\,.
\eea

\paragraph{Spinor brackets:}
We define spinor brackets such as $\dab 1 2 \dak \!= \dab p_1 |^a | p_2 \dak_a $ as the obvious inner-product of the 4-component Dirac spinors. These products are automatically antisymmetric $\dab 1 2 \dak = - \dab  2  1\dak$. If  we choose the same reference vector $q$ for the two momenta, then $\dab 1 2 \dak = \< 1^\perp 2^\perp\>$. The symmetric product $\dab 1 2 \dsk  = \dsb 2 1 \dak$ is non-vanishing for massive
fermions, but vanishes in the massless limit.

The spinor products are particularly simple when \emph{all reference vectors $q_i$ are equal}:
\bea
   \dab ij \dak = \< i^\perp j^\perp\> \, ,~~~~~~
   \dsb ij \dsk = [ i^\perp j^\perp] \, ,~~~~~~
     \dab i j \dsk
   \,=\,
       i\, m_i \frac{[ j^\perp q ]}{[ i^\perp q ]}
      - i\, m_j \frac{\< i^\perp q \>}{\< j^\perp q\>} \, .
\eea
Note also that $\dab i q \dsk = \dab q j \dsk = 0$. For further details, see appendix \ref{sec:SpinorHelicityFormalism}.

\paragraph{Massive vector bosons:}
As for massless vector
bosons, we can write the polarization vectors of the massive vector bosons in terms of the 4-component spinors:
\bea
  \nonumber
    \epsilon_-^{\dot\alpha\beta}&=&\frac{\sqrt{2}|p^\perp\>^{\dot\alpha} [q|^\beta}{[ p^\perp q]}
    ~~\,\qquad \Leftrightarrow\qquad
  \slashed{\eps}_- ~=~
  \frac{\sqrt{2}}{\dsb p q \dsk}
  \Big( |p\dak \dsbB q| - |q\dsk \dabB p| \Big)  \, ,
  \\[1mm] \label{theEps}
    \epsilon_+^{\dot\alpha\beta}&=&\frac{\sqrt{2}|q\>^{\dot\alpha} [p^\perp|^\beta}{\< p^\perp q\>}
    ~~\,\qquad \Leftrightarrow\qquad
  \slashed{\eps}_+ ~=~
  \frac{\sqrt{2}}{\dab p q \dak}
  \Big( |q\dak \dsbB p| - |p\dsk \dabB q| \Big) \, ,\\[1mm]
  \slashed{\epsilon}_0 &=&
   \frac{1}{m} \Big( \slashed{p}^\perp  - \frac{m^2}{\<q|p|q]} \slashed{q} \Big)
   ~=~\frac{1}{m} \Big( \slashed{p} - \frac{2m^2}{\<q|p|q]} \slashed{q} \Big) \, ,
    \nonumber
\eea
where $|q\dak=(0,|q\>)$ and $|q\dsk=(|q],0)$ are the usual massless spinors written in 4-component form. Together with $p_\mu/m$, the polarizations form a properly normalized basis of four-vectors.

Unlike in the massless case, the helicity amplitudes with massive particles depend explicitly on the reference vector $q$; this is simply because of the way we decompose the spin states in terms of the frame-dependent notion of $q$-helicity.

\subsection{\emph{q}-helicity little-group scaling}
\label{sec:lgs}
For the moment, let us choose different reference vectors $q_i$ for each external particle. Suppose we scale $q_i$ as
\bea
   \label{littlegrp}
   |q_i\> \to t_i^{-1} |q_i\> \,,
   ~~~~~~~
   |q_i] \to t_i\, |q_i] \,.
\eea
This implies
\bea
   |p_i^\perp\> \to t_i\, |p_i^\perp\> \,,
   ~~~~~~~
   |p_i^\perp] \to t_i^{-1} |p_i^\perp] \,,
\eea
and
\bea\label{doublebraLG}
  &&| i \dak \to t_i\, |i\dak\, ,~~~~~~
   | i \dsk \to t_i^{-1}\, |i\dsk\, ,~~~~~
  | i \dakB \to t_i\, |i\dakB\, ,~~~~~
     | i \dskB \to t_i^{-1}\, |i\dskB\, .~~~~~
\eea
So an outgoing Dirac fermion wavefunction scales as $t_i^{-2\tilde{h}_i}$ (incoming $t_i^{2\tilde{h}_i}$) where $\tilde{h}_i = \pm \tfrac{1}{2}$ is the $q$-helicity. It is clear from \reef{theEps} that the polarizations of outgoing vector bosons
also scale as $t_i^{-2\tilde{h}_i}$:
\bea
  \slashed{\eps}_{i\pm} ~\to~ t_i^{\mp2}\, \slashed{\eps}_{i\pm}
  \, ,~~~~~~~~~~~
  \slashed{\eps}_{i;0}~ \to~ t_i^{0}\, \slashed{\eps}_{i;0}\,.
\eea
Propagators do not scale. We conclude that under the scaling \reef{littlegrp} an on-shell amplitude with outgoing particles only scales as
\bea\label{lgsAn}
  A_n  \to t_{i}^{-2\tilde{h}_i} \, A_n \,. \label{phs}
\eea
We refer to this scaling as {\em $q$-helicity little-group scaling}.

In the following subsections, we will take all reference vectors to be equal, $q_i = q$, and all particles to be outgoing. From~(\ref{phs}), we conclude
\bea
   A_n \big(\{t^{-1} |q\>, t\, |q]\}\big)
   ~=~
   t^{-2 \sum_{i} \tilde{h}_i}
   \, A_n \big(\{|q\>, |q]\}\big)\,.
   \label{littlegrp2}
\eea
This is the massive equivalent of the little-group scaling we exploited to study the large-$z$ behavior of the all-line shift in section \ref{AllLineShifts}.

\subsection{All-line shifts in the massive spinor-helicity formalism}
\label{sec:massshift}
There is a natural implementation of the all-line shift \reef{massiveshift} in the framework of the massive spinor-helicity formalism of section~\ref{s:massSH}. First note that the spinors $|p^\perp]$ and $|p^\perp\>$ in~(\ref{pperps}) contain a normalization $1/\sqrt{\<q|p|q]}$. Clearly we want to avoid shifting $\<q|p|q]$. This is achieved by choosing all reference vectors equal, $q_i=q$, and by setting
\begin{equation}
    |X]~\equiv~ |q]\,.
\end{equation}
With this choice, the massless projections $p_i^\perp$ shift exactly as massless external momenta shift under the conventional anti-holomorphic shifts:
\begin{equation}
|\hat p_i^\perp\>~=~| p_i^\perp\>\, ,\qquad
|\hat p_i^\perp]~=~|p_i^\perp]+ z \,
 b_i |X]\, ,
\end{equation}
where  $b_i = \alpha_p \,d_i$ with $\alpha_p\!=\!\sqrt{\<q|p_i|X]}\!=\!\<q \,p^\perp \>\!=\!  [ p^\perp X]$.
The Dirac spinors $|p\dsk$ and $|p\dak$ shift in analogy with their massless counterparts:
\begin{equation}
    |\hat p\dak= |p\dak \,,\qquad |\hat p\dsk=|p\dsk+z \,
    b_i |X\dsk \,,~~~~~
    |X\dsk \equiv \begin{pmatrix} |X]\\ 0 \end{pmatrix}\,.
\end{equation}
Shifted angle and square spinor products have the same $z$-dependence as in the massless case,
\begin{equation}\label{spinorz}
    \dab\hat p_i\,\hat p_j\dak ~=~ \dab p_i\, p_j\dak \,,\quad\dsb\hat p_i\,\hat p_j\dsk ~=~ \dsb p_i\, p_j\dsk +
     z \big( b_j[p_i^\perp X]- b_i[p_j^\perp X] \big)\,,
\end{equation}
while mixed angle/square spinor products remain unshifted:
\begin{equation}
    \dsb \hat{p}_i \hat{p}_j \dak ~=~ \dsb p_i p_j \dak\,.
\end{equation}
The polarization vectors~(\ref{theEps}) shift as follows:
\bea
   \hat{\slashed{\eps}}_{i\,-} = \slashed{\eps}_{i\,-} \, ,~~~
  \hat{\slashed{\eps}}_{i\,+} = \slashed{\eps}_{i\,+} + z\,b_i \,\frac{\sqrt{2}}{\< p_i^\perp q \>}
  \Big( |q\> [X| - |X] \< q| \Big) \, ,~~~
  \hat{\slashed{\epsilon}}_{i\,0}=  \slashed{\epsilon}_{i\,0}
  + z\,  \frac{\slashed{r}_i }{m}  \, ,
  \label{shiftEps}
\eea
where $r_i$ is the momentum shift of line $i$, as defined in \reef{eq:genericalMomShift}.  The behavior of the transverse polarizations $\hat{\slashed{\eps}}_{i\,\pm}$ is just like for massless vectors whose reference spinors are proportional to the $|X]$ of the shift.

\subsection{Large-$z$ behavior under massive all-line shifts}

Under the anti-holomorphic all-line shift discussed in sections \ref{sec:mshift} and \ref{sec:massshift}, the amplitudes with massive particles among the external states behave as
\bea\label{massivelargez}
\boxed{\phantom{\Biggl(}
   \hat A_n(z) \to z^{\lfloor s\rfloor}\,~\text{(or better)\,,~~~~as}~~~z \to \infty\,,\qquad \text{with }~\,
   2s =4-n -c +\sum_i \tilde{h}_i \, .
   ~}
\eea
This is very similar to (\ref{falloff}) for the massless case, but with helicity replaced by $q$-helicity. As before, $c$ is the (lowest) mass dimension of the product of couplings in the Feynman diagrams for the process under consideration. As we show with the examples in the next section, processes with massive particles can have half-integer $s$, but the large-$z$ behavior has to have an integer exponent since the amplitude is a rational function. We argue below that the $\lfloor s\rfloor \equiv \mbox{floor}(s)$ that appears in~(\ref{massivelargez}) is indeed the appropriate integer bound on the exponent of the large-$z$ falloff.

To prove \reef{massivelargez}, we need to consider how
 the amplitude depends on angle/square spinors and momenta, and how the various brackets behave for large $z$ under the all-line shift.
 For now, we assume that no longitudinal gauge bosons are among the external states, and we will treat amplitudes with such particles separately at the end.
A generic term in the amplitude consists of
\begin{enumerate}
\item[1)] propagators $1/(P_I^2+m_I^2)$,
\item[2)] dot products of polarizations and momenta, $p_i\cdot p_j$, $\eps_i\cdot p_j$ and $\eps_i\cdot \eps_j$, and
\item[3)] spinor wave functions contracted into spinor products such as $\dab i | f(p,m,\eps) | j \dsk$, $\dab i | f(p,m,\eps) | j \dak$, and $\dsb i | f(p,m,\eps) | j \dsk$ with $f$ denoting polynomials in $\slashed{p}$, $\slashed{\eps}$, and $m$.
\end{enumerate}
 The dot products~2) and spinor products~3) can all be converted to products and sums of elementary anti-symmetric and symmetric spinor products so that the
 amplitude takes the schematic form\footnote{We do \emph{not} expand out the four-component spinor brackets according to their definitions.
In order to be able to use $q$-helicity little-group scaling in the following argument, it is important to use the spinor definitions~(\ref{doublebracketdef}) for both massive {\em and massless} particles. For example if $p_i^2=0$, we have
 $|i\dak=\bigl({}^{\,0\,}_{|i^\perp\>}\bigr)=t\bigl({}^{\,0\,}_{|i\>}\bigr)$
 and $|i\dsk=\bigl({}_{\,0\,}^{|i^\perp]}\bigr)=t^{-1}\bigl({}_{\,0\,}^{|i]}\bigr)$, with $t=\sqrt{[Xi]/\<iq\>}$.
}
\bea
  \label{formAn}
  A_n =  \sum g\,
  \frac{\dab . . \dak^{a_n}\dsb . . \dsk^{s_n}\dab . . \dsk^{t_n}
    \dab \, .\, q \dak^{\tilde{a}_n}
    \dsb\, .\, X \dsk^{\tilde{s}_n}
    \, m_.^{\m_n}
    }
    {
 (P_I^2+m_I^2)^{l_n}
    \dab \, .\, q \dak^{\tilde{a}_d}\dsb\, .\, X \dsk^{\tilde{s}_d}
    } \, .
\eea
Here $.$'s denote the external lines and $|q\dak$ and $|X\dsk$ are the reference spinors of the spinor-helicity formalism. For simplicity, we have used crossing symmetry to convert any in-going angle/square spinors to outgoing ones.
The Feynman rules show that the only factors that appear in the denominator are the
 $(P_I^2+m_I^2)$
of the propagators (our schematic notation includes massless propagators), and the antisymmetric brackets
$\dab \, .\, q \dak$ and $\dsb\, .\, X \dsk$. Spinor brackets involving $| q \dak$ or
 $|q \dsk =|X \dsk $ arise \emph{only} from the polarization vectors.  Since the expressions for the polarizations in \reef{theEps} are homogeneous in $|q \dak $ and  $|X \dsk$ respectively, the number of spinor brackets with $| q \dak$ and $| X \dsk$ appearing in the numerator of \reef{formAn} must be the same as in the denominator; hence $\tilde{a}_n=\tilde{a}_d$ and $\tilde{s}_n =\tilde{s}_d$.

Consider now any term in the sum~(\ref{formAn}), characterized by the mass dimension $c$ of its product of couplings $g$ and the non-negative integers $a_n$, $s_n$, $t_n$, $\tilde{a}_n$, $\tilde{s}_n$, $l_n$, and $\mu_n$.\footnote{We emphasize that explicit factors of masses $m$ that appear in the Feynman rules for interaction vertices (as is the case, for example, for Higgs bosons or massive gauge bosons) should be considered part of the product of couplings $g$ for an optimal estimate of the large-$z$ falloff. These masses then contribute to $c$, not to $\mu_n$.}
Every spinor bracket has mass dimension 1, so dimensional analysis gives
\bea
   \label{M1}
   a_n +s_n +t_n+\m_n  - 2l_n
   ~=~  4- n -c  \, ,
\eea
because the full amplitude \reef{formAn} must have mass dimension $4-n$. The \emph{q}-helicity little-group scaling~(\ref{doublebraLG}) leaves the spinor products $\dab ..  \dsk$, $\dab . q \dak$, and   $\dsb .  X \dsk$ invariant.
It is then obvious from \reef{formAn}  that
 the term scales as $t^{2(a_n-s_n)}$; on the other hand, \reef{lgsAn} requires a homogeneous  scaling $t^{- 2\sum_i\tilde{h}_i}$. We conclude that
$a_n -s_n$ must be the same for every term, namely
\bea
   \label{M2}
  a_n -s_n = - \sum_i\tilde{h}_i .
\eea
This is now combined with \reef{M1} to give
\bea
   \label{M3}
  2(s_n-l_n) ~=~  4- n -c
  -t_n-\m_n+\sum_i \tilde{h}_i\, .
\eea

Under an all-line shift, all propagators and all square brackets $\dsb . . \dsk$
 go as $\sim z$ for
large $z$. No other brackets shift; in particular, $\dsb\, .\, X \dsk$ is unshifted.  Each terms in \reef{formAn} therefore behaves as $z^{s_n-l_n}$
for large $z$; hence, the large-$z$ behavior of the full amplitude \reef{formAn} is determined by the terms with the largest value of $s_n-l_n$. As $t_n \ge 0$,  \reef{M3} allows us to conclude that
\bea\label{M4}
  2(s_n-l_n) ~=~  4- n -c -t_n-\m_n   ~\le~  4- n -c + \sum_i \tilde{h}_i \, .
\eea
Since each term goes as $\sim z^{s_n-l_n}$ for large $z$,
 and since $s_n-l_n$ must be an integer,~(\ref{M4}) proves the claim \reef{massivelargez} in the absence of longitudinal vector bosons among the external states.

\paragraph{Longitudinal gauge bosons:} Let us now include longitudinal gauge bosons among the external states.
First, let us note that the polarization vector $\epsilon_0$ in~(\ref{theEps}) is not the same as the longitudinal polarization $\epsilon_L$ in the conventional helicity basis (see~(\ref{usualeps}) for details). However, its leading contribution at large momenta (and in particular at large $z$) coincides with the leading contribution of $\epsilon_L$. To leading order in large $z$, we apply the equivalence theorem to the longitudinal gauge boson and replace it by its associated Goldstone boson $\Phi$.\footnote{See~\cite{Boels:2010mj} for a related application of the equivalence theorem in the context of BCFW shifts.}
We find
\begin{equation}\label{eq:scalarEquivTheorem}
(\hat \epsilon_0)_\mu\<\cdots \hat A^\mu \cdots\>~\sim~ (\hat\epsilon_L)_\mu\<\cdots \hat A^\mu \cdots\>~\sim~ \<\cdots \hat\Phi \cdots\>  \qquad\text{(leading order in $z$)}\,.
\end{equation}
Carrying out this replacement for each external gauge boson with polarization $\epsilon_0$, we can treat all of these particles as scalars (with $q$-helicity $\tilde{h}=0$) for the purpose of analyzing their large-$z$ behavior.
Note, however, that is crucial to perform the expansion~(\ref{formAn}) of the amplitude {\em after} replacing longitudinal gauge bosons by their corresponding scalars. The terms in the expansion (in particular the mass dimensions $c$ of their couplings $g$) will be different than for external gauge bosons because the Goldstone-bosons have different Feynman rules. We will see this in the example of $Z$-boson scattering in section \ref{sec:Z}.

In summary, we conclude that~(\ref{massivelargez}) holds for general amplitudes, as long as we determine the mass dimension $c$ of the couplings after replacing external longitudinal gauge bosons by their corresponding Goldstone boson $\Phi$. In particular, we have $c\geq 0$ for power-counting renormalizable theories, and thus all amplitudes with $n>4$ external legs are constructible in such models.

The rest of this section is dedicated to examples which illustrate the massive spinor-helicity formalism, the large-$z$ behavior, and how recursion relations work for massive particles. First we consider simple 4-point amplitudes in Yukawa theory (section \ref{sec:ExampleMassiveYukawaTheory}) and then amplitudes of $Z$ bosons in the electroweak theory (section \ref{sec:Z}).

\subsection{Examples: Yukawa theory with massive fermions}\label{sec:ExampleMassiveYukawaTheory}
Consider a real scalar field $\phi$ with mass $m_\phi$ coupled to Dirac fermions  $\Psi$  and $\overline{\Psi}$  with mass $m_e$ through the Yukawa interaction
$\lambda\, \phi \overline{\Psi} \Psi$.  We refer to the Dirac particles and antiparticles as electrons $e$ and positrons $\bar{e}$. The external states of the helicity amplitudes are specified in terms of the $q$-helicity basis introduced earlier in this section. All external states are outgoing and we write amplitudes as
\bea
\label{Astates}
A_\text{states}^\text{$q$-helicity}.
\eea
For example, $A_{e\,\,\bar{e}\,\, \phi\,\,\phi}^{- -\, 0 \,\,0}$ denotes the scattering amplitude whose outgoing external states are an electron and positron, both with $q$-helicity $\tilde{h} = -1/2$, and two scalars $\phi$ with $\tilde{h}=0$.

Start with the {\bf 3-point amplitudes}. Following the Feynman rules for the fermion wave functions \reef{Dfeyn}, we have
\bea
  A_{e\,\,\bar{e}\,\,\phi}^{--\,0} = \lambda\, \dab 1 2\dak \, ,~~~~
  A_{e\,\,\bar{e}\,\,\phi}^{++\,0} = \lambda\,\dsb 1 2 \dsk\, ,~~~~
  A_{e\,\,\bar{e}\,\,\phi}^{-+\,0} = \lambda\, \dab 1 2 \dsk\, ,~~~~
  A_{e\,\,\bar{e}\,\,\phi}^{+-\,0}  = \lambda\, \dsb 1 2 \dak\,.
  \label{eq:Aempmphi}
\eea
The last two amplitudes are proportional to the fermion mass
$m_e$ and vanish in the massless limit.

\vspace{2mm}

\noindent {\bf 4-point electron-positron scattering:}\\
Consider the process with two outgoing electrons and two outgoing  positrons. Let us first take all four $q$-helicities to be $\tilde{h}=-1/2$. There are two Feynman diagrams and the Feynman rules
\reef{Dfeyn}
 directly give
\bea
\label{Aeeee1}
A_{e\,\, \bar{e}\,\, e\,\, \bar{e}}^{----} = \lambda^2 \frac{\dab 1 2 \dak \dab 3 4 \dak}{P_{12}^2 + m_{\phi}^2} ~-~ (1 \leftrightarrow 3).
\eea
The relative minus sign arises from the exchange of identical fermions.
Let us now apply the all-line shift to \reef{Aeeee1}.
The angle brackets $\dab .. \dak$ are unshifted, and we get a $1/z$-falloff from the propagator in each diagram.
This matches exactly the result \reef{massivelargez} of
our general analysis: since the coupling $\lambda$ is dimensionless, we have  $2s = 4-n + \sum \tilde{h}_i = -2$, giving
 a falloff $z^{s} = z^{-1}$. The all-line shift recursion relations are therefore valid for this amplitude; the recursion diagrams are just
$A_{e\,\, \bar{e}\,\, \phi}^{--0} \times \frac{1}{p^2 + m_\phi^2} \times A_{e\,\, \bar{e}\,\, \phi}^{--0}$.

Next let us flip the $q$-helicity of one of the positrons.
Then the Feynman rules give
\bea
A_{e\,\, \bar{e}\,\, e\,\, \bar{e}}^{---+} = \lambda^2 \frac{\dab 1 2 \dak \dab 3 4 \dsk}{P_{12}^2 + m_{\phi}^2} ~-~ (1 \leftrightarrow 3).
\eea
Under an all-line shift (recalling that $\dab i j \dsk$ is unshifted) this amplitude also has a $1/z$-falloff,
and this agrees with \reef{massivelargez}, where $\sum \tilde{h}_i = -1$ gives $z^{\lfloor s \rfloor} = z^{\lfloor -1/2 \rfloor} = 1/z$.
 The recursive calculation of this amplitude is trivial because the 3-point subamplitudes $A_{e\, \bar{e}\, \phi}^{-\pm 0}$ are unaffected by the shift.

Finally, consider electrons and positrons of both $q$-helicities.
The Feynman rules give
\bea
A_{e\, \,\bar{e}\,\, e\,\, \bar{e}}^{--++} =
 \lambda^2 \left(\frac{\dab 1 2 \dak \dsb 3 4 \dsk}{P_{12}^2 + m_{\phi}^2} - \frac{\dsb 3 2 \dak \dab 1 4 \dsk}{P_{23}^2 + m_{\phi}^2}\right).
\eea
Under the all-line shift, the first term is $\mathcal{O}(1)$ while the second is $\mathcal{O}(1/z)$.  This matches the
result
\reef{massivelargez} with
$\sum \tilde{h} = 0$. Thus this amplitude is not all-line-shift constructible.

\vspace{2mm}
\noindent {\bf 4-point electron-positron-$\phi$-$\phi$ scattering:}\\
The all-line shift recursion relation is less trivial for electron-positron-$\phi$-$\phi$ scattering.
There are two Feynman diagrams, related by exchanging the scalar lines $3 \lra 4$, so we have
\bea\label{eq:Aempmphiphi}
A_{e\,\, \bar{e}\,\, \phi\,\, \phi}^{--\,0\,\,0}
 = \lambda^2 \frac{\dab 1| \slashed{p}_{23} + i m_{e}|2  \dak}{P_{23}^2+m_{e}^2} + (3 \leftrightarrow 4).
\eea
The numerator does not shift under the all-line shift, so the amplitude has a $1/z$-falloff, as expected from $ z^{\lfloor s \rfloor} =z^{\lfloor -1/2 \rfloor} =z^{-1}$.

Now we construct the amplitude \reef{eq:Aempmphiphi} using the all-line shift recursion relations. We present the full details in order to illustrate the use of the 4-component massive spinor-helicity formalism. The 2-3 pole recursion diagram
has two contributions since we have to sum over the helicity of the internal fermion line; taking all lines on the subamplitudes to be
outgoing, we have
\begin{equation}
\begin{split}
  A_{e\,\, \bar{e}\,\, \phi\,\, \phi}^{--\,0\,\,0}
  &=
  \sum_{\pm}
  {A}^{-\mp
  \, 0}_{e\,\,\bar{e}\,\,\phi}(\hat{p}_1,\hat{p}_{23},\hat{p}_4)\,
  \frac{1}{P_{23}^2+m_{e}^2}\,
  {A}^{\pm
  -\,0}_{e\,\,\bar{e}\,\,\phi}(-\hat{p}_{23},\hat{p}_2,\hat{p}_3)\,
  + (3 \leftrightarrow 4)\\[1mm]
  &= \frac{\dab 1 \, \hat{p}_{23} \dak  \dsb (-\hat{p}_{23}) \, 2 \dak
     + \dab 1 \, \hat{p}_{23} \dsk  \dab (-\hat{p}_{23}) \, 2 \dak  }
 {P_{23}^2+m_{e}^2}
 + (3 \leftrightarrow 4)\\[1mm]
  &= \frac{\dab 1 \, \hat{p}_{23} \dak  \dsbB \hat{p}_{23} \, 2 \dak
     - \dab 1 \, \hat{p}_{23} \dsk  \dabB \hat{p}_{23} \, 2 \dak  }
 {P_{23}^2+m_{e}^2}
 + (3 \leftrightarrow 4)\\[1mm]
  &= \frac{\dab 1 | \,\hat{\slashed{p}}_{23} + i m_e
  | 2 \dak }
 {P_{23}^2+m_{e}^2}+ (3 \leftrightarrow 4)\,.
 \label{resD23}
\end{split}
\end{equation}
In the second line we used that the angle brackets are unshifted, in the third line we used the rule for crossing symmetry, and in the final
step
we applied the completeness relation \reef{complete1}.
Now in \reef{resD23} the `hat' on
$\hat{\slashed{p}}_{23}$ can be removed since the shifted part gives terms proportional to $\dsb X 2 \dak$, which vanish because $|q] = |X]$. This reproduces the result \reef{eq:Aempmphiphi}. We would have arrived at the result more directly by taking the
internal states to be incoming fermions;  however,
we found the manipulations above to be a useful illustration of the internal consistency of the formalism.

\subsection{Examples: $Z$-boson scattering}
\label{sec:Z}

Now let us illustrate the all-line-shift constructibility for massive $Z$-boson scattering. In the electro-weak model, there is a 3-vertex interaction between two $Z$-bosons and the Higgs $h$. There is no 4-vertex with four $Z$-bosons, so for an amplitude with four external $Z$-bosons the Feynman diagrams  are all
 Higgs-boson-exchange diagrams.

\vspace{2mm}
\noindent {\bf 3-point amplitudes:}\\
The $ZZh$ 3-vertex is $(g\, m_Z/c_{W})\, \eta^{\mu\nu}$, where $g$ is the $SU(2)$ coupling constant, $m_Z$ is the $Z$-boson mass and $c_{W}$ is the cosine of the weak mixing angle. Dotting in the $Z$-polarizations \reef{theEps} we find
\bea
A_{ZZh}^{-+0} = \frac{g\, m_Z}{c_{W}}\frac{\dabB 1 q \dak \dsbB 2 q \dsk}{\dsb 1 q \dsk \dab 2 q\dak} \, ,~~~~
A_{ZZh}^{\,0\,0\,0} = \frac{g}{2\, m_Z c_{W}}\left(\!
\dab1 2 \dak\dsb1 2\dsk
\!+\! m_Z^2
\frac{\<q|1|q]^2\!+\!\<q|2|q]^2}{\<q|1|q]\, \<q|2|q]}\right)\,,
\eea
while $A_{ZZh}^{\pm\pm0}=0$.
We label the amplitude with $q$-helicity superscripts as in \reef{Astates}. We have used identities such as $\dab 1 q \dakB = \dabB 1 q \dak$ and
 $2 p^\perp \cdot q = - \<q|p|q]$  to simplify the results.

\vspace{2mm}
\noindent {\bf 4-point amplitudes:}\\
In the $q$-helicity basis, all amplitudes with
 $\sum \tilde{h}_i\neq 0$ vanish:
 $A_{ZZZZ}^{++++} = A_{ZZZZ}^{+++-} = A_{ZZZZ}^{+++\,0} = ... = 0$. For the 4-point amplitude $A_{ZZZZ}^{-+-+}$
a Feynman diagram calculation gives
\bea
\label{ZZZZ1}
A_{ZZZZ}^{-+-+} &=& \frac{g^2 m_Z^2}{c_{W}^2}\,
\frac{
(\epsilon_{1-} \cdot \epsilon_{2+})\,
(\epsilon_{3-} \cdot \epsilon_{4+})}
{P_{12}^2+m_h^2} ~~+~ (1 \leftrightarrow 3)\nonumber\\
&=& \frac{g^2 m_Z^2}{c_{W}^2}\,\,\frac{\dabB 1 q \dak \dsbB 2 q \dsk}{\dsb 1 q \dsk \dab 2 q\dak}\times\frac{1}{P_{12}^2+m_h^2}\times\frac{\dabB 3 q \dak \dsbB 4 q \dsk}{\dsb 3 q \dsk \dab 4 q\dak} ~~+~ (1 \leftrightarrow 3),
\eea
where the two contributions come from $s$- and $u$-channel Higgs-boson exchange.  Under an all-line shift with $|X]=|q]$ this amplitude
 goes as  $1/z$ since double-brackets of the form $\dsb i q \dsk$ are invariant under this shift. This matches the result
\reef{massivelargez} with $n=4$, $c=2$ (the product of couplings is proportional to $m_Z^2$ due to electro-weak symmetry breaking),
and $\sum \tilde{h}_i=0$. The good large-$z$ falloff means that the amplitude \reef{ZZZZ1} can be constructed using the all-line recursion relations:
 the recursion diagrams are simply $A_{ZZh}^{-+0} \times\frac{1}{p^2+m_h^2} \times A_{ZZh}^{-+0}$.

It is particularly instructive to study the scattering of four longitudinal $Z$ bosons.  In this case, there are contributions to the Feynman diagram calculation from $s$-, $t$- and $u$-channel Higgs-boson exchange:
\bea
A_{ZZZZ}^{\,0\,0\,0\,0} \!\!\!&=&\!\!\!
 \frac{g^2 m_Z^2} {c_{W}^2}
\frac{
(\epsilon_{1;0}\cdot\epsilon_{2;0}) \,
( \epsilon_{3;0} \cdot\epsilon_{4;0})}{P_{12}^2+m_h^2}
 ~~+~ (1 \leftrightarrow 3)  ~+~ (1 \leftrightarrow 4)\nonumber\\[.3ex]
&=& \frac{g^2}{4\,m_Z^2c_{W}^2}\left(
 \dab1 2\dak\dsb1 2\dsk
+ m_Z^2
 \frac{\<q|1|q]^2+\<q|2|q]^2}{\<q|1|q]\, \<q|2|q]}
\right)
\times\frac{1}{P_{12}^2+m_h^2}\nonumber\\&&~~~\times
\left(
 \dab 34 \dak\dsb 34\dsk
+ m_Z^2
 \frac{\<q|3|q]^2+\<q|4|q]^2}{\<q|3|q]\, \<q|4|q]}
\right)
 ~\,
 +\, (1 \leftrightarrow 3)  \,+\, (1 \leftrightarrow 4).\label{A0000}
\eea
Note that since $\dsb \hat i \hat j\dsk\sim z$,
each diagram in this amplitude goes as $\mathcal{O}(z)$ under an all-line shift.  For the sum of all terms, however, there is a cancellation of the leading order piece and the overall scaling is actually $\mathcal{O}(1)$. To see this, first note that  to leading order the propagator gives
\begin{equation}
    \hat P_{12}^2+m_h^2\,\sim~ \dab\hat 1 \hat 2\dak\dsb\hat 1 \hat 2\dsk\,\,\sim~ z (d_1-d_2) [X|\slashed{p}_2\,\slashed{p}_1|X]\,.
\end{equation}
Since
the leading  term in the numerator of~(\ref{A0000}) comes from
$\bigl(\dab\hat 1 \hat 2\dak\dsb\hat 1 \hat 2\dsk\bigr)^2$, it follows that
\begin{equation}\label{leadingAZZZZ}
    \hat A_{ZZZZ}^{\,0\,0\,0\,0} \,=\, z\,\frac{g^2}{4m_Z^2 c_W^2}\Bigl[ (d_1-d_2) [X|\slashed{p}_2\,\slashed{p}_1|X]+(1 \leftrightarrow 3)  \,+\, (1 \leftrightarrow 4)\Bigr]+O(z^0)\,.
\end{equation}
Using  momentum conservation $\sum_ip_i=0$ and the constraint~(\ref{massiveshift}) on the shift parameters $d_i$, it is easy to see that the term in brackets in~(\ref{leadingAZZZZ}) vanishes. Therefore,
\begin{equation}
    \hat A_{ZZZZ}^{\,0\,0\,0\,0}~\sim~ z^0\,.
\end{equation}
This is the Goldstone boson equivalence theorem in action!
According to our prescription
\reef{eq:scalarEquivTheorem}, all
longitudinal
vector
bosons must be replaced by Goldstone bosons to determine the large-$z$ scaling. Then~(\ref{massivelargez}) predicts the large-$z$ behavior of this amplitude correctly. Indeed, the large-$z$ limit is dominated by the diagram
with
a
four-Goldstone contact interaction, which scales as $\mathcal{O}(1)$.

\setcounter{equation}{0}
\section{When are tree amplitudes on-shell constructible?}\label{sec:DiscussionOfConstructibility}
Why are some tree amplitudes constructible via on-shell recursion relations while others are not?  Here, we propose a simple physical interpretation of why the all-line shifts fail for certain classes of amplitudes. We first describe some general ideas, and then apply them in the context of all-line shifts.

\subsection{General ideas}
Consider a theory described by a local Lagrangian. An on-shell tree amplitude $A_n$ with $n$ external states depends only on interaction vertices with
$m<n$
fields. However, if $A_n$ can be computed by on-shell recursion relations, then it has an expression in terms of lower-point on-shell
 amplitudes. In particular, there is a way to determine $A_n$ without explicit knowledge of any local $n$-point
 contact-term interactions.

 Yang-Mills theory, for example, has 3- and 4-point interaction vertices.  The fact that there are valid recursion relations (such as BCFW) for all Yang-Mills
amplitudes with $n>3$ external lines means that all amplitudes are completely determined by the basic 3-point vertex. The irrelevance of the 4-point vertex for the \emph{on-shell} amplitudes is not surprising: it has to be included with the 3-point interaction only to make the \emph{off-shell} Lagrangian gauge invariant.

Generalizing the Yang-Mills example, we say that an $n$-point interaction $Y$ in the local Lagrangian  is a  \emph{dependent interaction}, if $Y$ is completely determined by lower-point interactions (for example through gauge invariance or symmetries, such as supersymmetry).
On the other hand, we refer to $Y$ as an \emph{independent interaction} if the Lagrangian is gauge-invariant and respects all imposed symmetries
without the inclusion of $Y$. The basic idea is that dependent
 $n$-point interactions should not be required as input for on-shell  amplitudes, while the information from independent
 $n$-point interactions must be supplied directly as it cannot be obtained recursively from on-shell amplitudes with less than $n$ external states.

Let us use scalar QED to illustrate this idea.
 The kinetic term $|D\phi|^2$ gives rise to  3- and 4-point interactions
$A_\m \, \phi\,  \partial^\m \bar{\phi}$ and $\phi\,\bar{\phi} A_\m A^\m$. The 4-point
interaction
is required by gauge invariance, so it is dependent. Indeed the process $\< \phi\,\bar{\phi} + - \>$ can be calculated without the use of the 4-vertex, for example using BCFW recursion relations.
How about the 4-scalar process $\< \phi\,\bar{\phi}\, \phi\,\bar{\phi} \>$?
It has pole term contributions from the 3-vertices $A_\m \, \phi\,  \partial^\m \bar{\phi}$, but if the theory has a 4-point contact interaction $\lambda |\phi|^4$, then there is \emph{a priori} no mechanism for
the vertices of $|D\phi|^2$ to determine $\lambda$. Thus $\lambda |\phi|^4$ is
an
example of an independent interaction whose input is needed for 4-point amplitudes; but once the 4-point amplitudes are supplied in addition to the 3-point amplitudes, then 5- and higher-point amplitudes are calculable recursively, for example with all-line shift recursion relations.

 We expect constructibility properties of amplitudes to improve when extra symmetries are available.
This is nicely illustrated in theories with supersymmetry.
For example, the
$\cn=4$ SYM Lagrangian has a 4-scalar interaction  of the form $Y=\phi^{12} \phi^{23} \phi^{34} \phi^{14}$. The color-ordered tree amplitude with these four external scalars has no pole contributions and hence the only contribution is from $Y$, $\< \phi^{12} \phi^{23} \phi^{34} \phi^{14}\> = 1$.
 This amplitude cannot be calculated by standard BCFW recursion relations.  However, $Y$  is part of the \emph{unique}
 $\mathcal{N}=4$  supersymmetric completion of the basic Yang-Mills 3-vertex $A^2 \partial A$,  and hence $Y$ is a dependent interaction once we impose supersymmetry. We can only expect a recursive computation of the amplitude
 $\< \phi^{12} \phi^{23} \phi^{34} \phi^{14}\>$ from a complex shift that
 respects supersymmetry.
 The super-BCFW shift of~\cite{ArkaniHamed:2008gz,Brandhuber:2008pf} respects `super-momentum conservation'.
The
associated
recursion relations
allow one to compute
\emph{any} tree amplitude of $\cn=4$ SYM with $n\ge 4$ external legs from the basic 3-point superamplitude, which is fully determined by supersymmetry from the 3-vertex $A^2 \partial A$.

To summarize: interactions $Y$  that derive from lower-point interactions
 by gauge invariance or symmetries should not be needed as separate input in recursion relations that respect these symmetries. All on-shell recursion relations incorporate gauge invariance, so gauge-dependent interactions are not expected to provide independent input. This means that we can focus on the leading interaction
of
gauge-invariant operators,  such as the 3-point interaction of ${\rm tr}\,\phi\, F^2$, while the higher-point dependent interactions give constructible contributions. However, sometimes there can be ambiguities in the non-linear gauge completion of an operator.
We discuss examples of this and the consequences  for on-shell constructibility in the next section where we focus on the all-line shift recursion relations.

\subsection{Interpretation of all-line shift constructibility}
For anti-holomorphic all-line shift, the condition for large-$z$ falloff derived in section \ref{AllLineShifts} is
\bea\label{falloff2}
   \hat{A}_n(z) \to 0\,~\text{~~~as}~~~z \to \infty\,,~\quad\text{when }\quad 4-n -c +\sum_i h_i ~<~ 0\, .
\eea
Here $c$ is the (smallest) mass dimension of the product of couplings entering the calculation of
 $A_n$, and $\sum_i h_i$ denotes the sum of helicities of the external states.
The bound in \reef{falloff2} was derived utilizing only
dimensional analysis and little-group scaling, and therefore it applies very generally. In particular, it must take all possible local gauge-invariant interactions
 of dimension $c$ into account, whether or not the particular theory we have in mind contains such interactions.  We will exploit the following: if an amplitude
 $\hat{A}_n$ vanishes as $z \to \infty$, then  the Lagrangian cannot contain any independent $n$-point interactions that contribute to $A_n$.
 Conversely, if gauge-invariance does not fix the $n$-point interactions completely, then we cannot expect the all-line shift recursion relations to be
 valid. Therefore $\hat{A}_n$ cannot vanish as $z \to \infty$. We now use examples to illustrate this proposal for the interpretation of the bound in \reef{falloff2}.

Consider
 adding an operator  $\lambda\,D^{2q} F^m$
to the pure Yang-Mills Lagrangian.\footnote{Henceforth we suppress the trace structure as it does not play a role for our arguments.} The coupling $\lambda$ has mass dimension $4\!-\!2m\!-\!2q$ (see Table~\ref{tab:cValues}). We consider tree amplitudes  $\<  \dots  \>_{D^{2q}\!F^m}$ with a single insertion of this operator; clearly, such amplitudes only exist for $n \ge m$. Let us start with the   amplitudes $\<  -- \dots -  \>_{D^{2q}\!F^m}$ where all $n$ external states are negative helicity gluons.\footnote{
Such amplitudes can be non-vanishing only when the operator is not supersymmetric.}
We cannot expect amplitudes with $m$ external legs to be on-shell constructible
since the leading $m$-point interaction of $D^{2q} F^m$ contains independent information not available in pure Yang-Mills  theory. Should we then expect all-line shift recursion to
work
for any all-minus amplitude with $n>m$ external legs?
First consider $F^m$, the case with no additional covariant derivatives $(q=0)$.  Here the answer is yes:~the result~(\ref{falloff2}) implies constructibility for $n> m$. Since the higher-point gluon interactions in $F^m$ are completely determined by gauge-invariance, the $m$-point input is sufficient to compute all higher-point all-minus amplitudes recursively.

For $D^2F^m$, however, validity of the all-line shift recursion relations
 (\ref{falloff2})  requires $n> m+1$,
so the
$(m\!+\!1)$-point amplitudes  $\<  -- \dots -  \>_{D^{2}\!F^m}$ are not all-line-shift constructible.
Indeed, an independent gauge-invariant operator
$\lambda' \, F^{m+1}$ can be added to the Lagrangian. It has the same mass dimension as $D^2F^m$ and contributes a local term to the $(m+1)$-point gluon amplitudes. Due to the general nature of the argument that led to \reef{falloff2}, the result has to allow for the possible presence of $F^{m+1}$-interactions.
In other words, the gauge-invariant non-linear completion of $D^2F^m$ is ambiguous at the $(m+1)$-point level, and may contain an arbitrary linear combination of $F^{m+1}$ operators. This is why
the all-minus $n$-point amplitudes with one insertion of $D^2F^m$ are only constructible for  $n>m+1$; both the $m$ and $(m+1)$-point amplitudes need to be supplied as an input for the recursion relation to resolve the ambiguity.

For $D^{2q}F^m$, the result \reef{falloff2} with $c=4-2m-2q$ and $\sum_i h_i = -n$ shows that
\be\label{eq:conditionAllMinus}
\< - - \dots -- \>_{D^{2q}\!F^m}
~~~~\text{is constructible for }~~~
n> m+q \,.
\ee
 Again, we can understand this bound by exploring the possibilities for single insertions of independent gauge-invariant operators
 that can contribute to the amplitudes with $n=m+1,m+2,\ldots,m+q$ external legs.
Any gauge invariant operator that affects $\< - - \dots -- \>_{D^{2q}\!F^m}$
must be composed of field strengths $F$ and covariant derivatives $D$.
The ambiguity in the gauge-invariant interactions thus includes the set of operators
$D^{2q-2} F^{m+1}$, $\ldots$, $F^{m+q}$\,,
which all have the same mass dimension as $D^{2q} F^m$. Therefore, one cannot expect  all-line shift recursion  relations to be valid
for $n$-point all-minus amplitudes until $n>m+q$, exactly as \reef{falloff2} states.

Next, let us see what happens when we consider amplitudes with both positive- and negative-helicity gluons.
 This change leads to a
\emph{qualitatively different interpretation} of the bound on the validity of the all-line
  shift recursion relation. Let us start with $\< -- \dots - - +\>_{F^m}$, again with just a single insertion of the operator
 $\lambda\,  F^m$.
For this class of amplitudes $\sum_i h_i = 2-n$,  so  \reef{falloff2} shows that
\be
\label{nmq}
\< - - \dots -+ \>_{F^m}
~~~~\text{is constructible for }~~~
n> m+1 \,.
\ee
Thus to ensure all-line constructibility for $\< - - \dots -- + \>_{F^m}$,
 we  need one more external
line
  than for the all-minus amplitudes \reef{eq:conditionAllMinus}.
The reason is the following. The derivation of the condition \reef{nmq} relies only
on the {\em sum} of all helicities. For the all-minus amplitudes, the sum was $\sum_i h_i=-n$ and this uniquely identified the external states as negative helicity gluons. The sum
 $\sum_i h_i = 2-n$ of $\< - - \dots -+ \>_{F^m}$, however,  can be obtained by several different combinations of external states, for example $\< -- \dots - \phi \,\phi \>$  with some scalar field $\phi$.
The bound \reef{falloff2} must also be valid for this
 amplitude  --- this condition gives an upper bound on the worst-behaved
 amplitude in the entire  \emph{class} of amplitudes with $\sum_ih_i=2-n$.
Thus we  also  need to consider operators  that have the same coupling dimension as $F^m$ and  contain 2 scalars $\phi$ in additions to
vector fields.
A scalar has mass dimension 1, so a candidate independent operator is the
$(m\!+\!1)$-field operator
$\lambda'\, \phi^2 F^{m-1}$. It contributes to
the $(m\!+\!1)$-point amplitudes with $\sum_ih_i=2-n$,
  and
$\lambda'$ has the same mass dimension as $\lambda$, so
    $c=4-2m$.
Hence, we only expect recursion relations to be valid for $n> m+1$,
and this is in exact agreement with the bound \reef{nmq}.

We could continue by studying general N$^k$MHV amplitudes or multiple insertions of the operators ${D^{2q}\!F^m}$, but let us instead move on to another example from section \ref{s:examples}. In gluon-Higgs fusion (section \ref{sec:GluonHiggsFusion}), we considered the operator $Z= {\rm Re}(\phi)\, F^2+ {\rm Im}(\phi)\, F \tilde{F}$. We found that N$^k$MHV amplitudes with a \emph{single} insertion of $Z$ have large-$z$ falloff $z^{-k}$ under the all-line shift, so the all-line shift recursion relations (which in this case imply an MHV vertex expansion) are valid for any $k>0$ amplitude. Now consider multiple insertions of the operator $Z$, for example for the amplitude $\< -- \phi^{n-2}\>$ with $(n-2)$ insertions of $Z$. Then $c= 2-n$ and $\sum_i h_i =-2$, so \reef{falloff2} shows that the all-line shift deformed amplitude $\< -- \phi^{n-2}\>$ behaves as $z^0$
for large $z$. The ambiguity responsible for the failure of the all-line shift recursion relations for $\< -- \phi^{n-2}\>$ is clear: the amplitude can be affected by  the independent operator $\phi^{n-2} F^2$.

In our interpretation of the bound \reef{falloff2}, operators that can be removed by a field redefinition should not be considered as independent interactions.
Consider for example $\sqrt{-g}\,\phi^m R$, where $R$ is the Ricci scalar. A Weyl transformation reduces this operator to
$R$ without affecting the graviton amplitudes. The leading $(m\!+\!1)$-interaction involves two powers of the graviton momentum which in the
$(m\!+\!1)$-point  on-shell matrix element can only contract with each other or with the graviton polarization;
 the contribution vanishes in either case, so $\< \phi^m \pm \>_{\phi^m\!R} =0$.

As our last example, let us explain the behavior of the all-line shift for graviton amplitudes in pure Einstein gravity. We showed in section \ref{s:simplestEx} that large-$z$ falloff requires $n-3-2k <0$ for $n$-point N$^k$MHV amplitudes.  For simplicity, let us consider only the anti-MHV amplitudes $\<- \dots - - - ++\>$.
They have $k=n-4$, and are therefore constructible for $n>5$.
To interpret this bound, we want to identify an independent 2-derivative operator\footnote{In (super)gravity, we normalize all fields by the gravitational coupling $\kappa$, so
all 2-derivative interactions
 have the same coupling dimension as the interactions of the Einstein-Hilbert action.} whose leading 5-point interaction can contribute to the class of amplitudes with $\sum_i h_i =- 2(n-4)$. This helicity sum can be obtained from several combinations of external states. One option is four scalars and $n-4$ negative helicity gravitons, $\<- \dots - \phi^4\>$.
As discussed above, the operator $\sqrt{-g}\,\phi^4 R$ is not relevant since it can be removed by a field redefinition.
Next consider a sigma-model term
$\sqrt{-g}\,g^{\m\n} \phi_a \phi_b\, \partial_\m \phi_c \partial_\n \phi_d$. It can contribute, but only its 4-point interaction provides independent information. We have to consider a different set of external states to understand the bound $n>5$  from \reef{falloff}. Take two negative helicity gluons, three scalars and $(n\!-\!5)$ negative helicity gravitons: the sum of their helicities is $- 2(n-4)$. The 2-derivative operator $\sqrt{-g}\,\phi^3 F^2$ contributes an independent 5-point interaction to this class of amplitudes. This is the reason why all-line shift constructibility cannot start until
6-points for this class of amplitudes.

Supersymmetry did not feature in our above discussion of all-line shift constructibility. In fact, the all-line shift does not preserve supersymmetry  because it only leaves half of the super\-charges invariant. Therefore we cannot expect it to produce a falloff for amplitudes that are completely determined from the non-linear supersymmetric completion of lower-point interactions. The supersymmetry-preserving all-line {\em super}shift introduced in~\cite{Kiermaier:2009yu} is a natural candidate for this extension, and it would be interesting to generalize our analysis here to
super-all-line shift
recursion relations.

\section*{Acknowledgments}
We thank N.~Arkani-Hamed,
D.~Berenstein,
F.~Cachazo, C.~Berger, L.~Dixon,
J. Maldacena,
C.~Peng, A.~Pierce,
and
E.~Yao for useful
 discussions and suggestions.

HE is supported by NSF CAREER Grant PHY-0953232, and in part by the US Department of Energy under DOE grants DE-FG02-95ER40899 (Michigan) and DE-FG02-90ER40542 (IAS).
The research of MK is supported by NSF grant PHY-0756966. TC is supported in part by NSF CAREER Grant PHY-0743315.  TC wishes to thank the Institute of Advanced Study for hospitality during the completion of this work.

\def\theequation{\Alph{section}.\arabic{equation}}
\begin{appendix}

\setcounter{equation}{0}
\section{Review: Deriving the MHV vertex expansion in $\mathcal{N} \!=\! 4$ SYM}
\label{sec:allline2CSW}
We review here the steps needed to derived the MHV vertex expansion from the all-line shift recursion relations in $\mathcal{N}=4$ SYM theory \cite{Elvang:2008vz}.

{\it Step 1:}
The all-line shift recursion relation expresses an N$^k$MHV amplitude in terms of diagrams with two on-shell N$^{k_i}$MHV subamplitudes which have $k_1+k_2 = k-1$. The kinematics of the anti-holomorphic shift ensures that diagrams with a $k_i=-1$ vertex (3-point anti-MHV) vanish, and hence the subamplitudes have $k_i<k$.  Thus, if the relations are applied iteratively $k$ times, the amplitude will be expressed in terms of diagrams with MHV vertices only. This is not yet the MHV vertex expansion, but (as we will show in step 3) the diagrams can be resummed into MHV vertex diagrams.

\noindent {\it Step 2:}  The $k+1$ holomorphic MHV vertices of each all-line shift diagram depend on the shifted internal momenta $\hat{P}_{I_i} = |\hat{P}_{I_i}\rangle [\hat{P}_{I_i}|$ only through the angle brackets, $|\hat{P}_{I_i}\rangle = P_{I_i}|X]/[\hat{P}_{I_i} X]$.
Here $|X]$ is the reference spinor of the shift \reef{allline}.
The product of MHV vertices in the diagram is invariant under little-group scalings of the internal $\hat{P}_{I_i}$ spinors,
so all factors $[\hat{P}_{I_i} X]$ cancel. Therefore all $|\hat{P}_{I_i}\rangle$ can be replaced by $P_{I_i}|X]$.
This eliminates the details of the shift, and is exactly the CSW prescription \reef{CSWpres}.

\noindent {\it Step 3:} Each all-line shift diagram has one unshifted propagator $1/P_{I_i}^2$,
and the
$k\!-\!1$
other propagators
are evaluated at the particular value $z_i$  that takes $\hat{P}_{I_i}$ on-shell. For a given set of $k$ propagators, there are $k$ such diagrams, namely one for each choice of unshifted propagator $1/P_{I_i}^2$. The product of
$k\!+\!1$
MHV subamplitudes  is the same in these $k$ diagrams, since the vertices do not depend on $z_i$, as shown in step 2. This allows us to factor out the overall
MHV vertex dependence and sum the propagator factors.  The result simplifies due to a contour integral identity \cite{BjerrumBohr:2005jr,Elvang:2008vz}, and we find
\be
  \left(A_{(1)}^\text{MHV} \cdots A_{(k+1)}^\text{MHV}\right) \times   \sum_{i=1}^k \frac{1}{\hat{P}_{I_1}^2(z_i)\cdots P_{I_i}^2\cdots\hat{P}^2_{I_k}(z_i)}
  ~=~
   \frac{A_{(1)}^\text{MHV} \cdots A_{(k+1)}^\text{MHV}}
   {P_{I_1}^2 P_{I_2}^2 \cdots P_{I_k}^2} \, .
\ee
This is precisely the value of the MHV vertex diagram. The all-line shift recursion relations show that the amplitude is the sum of all such diagrams. This completes the derivation of the MHV vertex expansion from the all-line shift.

\setcounter{equation}{0}
\section{Spinor helicity formalism}\label{sec:SpinorHelicityFormalism}
We present here a self-contained outline of the spinor-helicity formalism used in this paper. In the main text we use a $q$-helicity basis defined in terms of an arbitrary null vector $q$. This differs from the more conventional helicity basis. We present the details of both bases and outline how to map between them.

We use a ``mostly-plus'' metric $\eta_{\mu\nu} = \text{diag}(-1,+1,+1,+1)$, and our Clifford algebra is $\{\g^\mu, \g^\nu \} = 2 \eta^{\mu\nu}$ with
\bea
  \g^\mu =
  \left(
     \begin{array}{cc}
        0 & \sigma^\mu \\
        \bar{\sigma}^\mu & 0
      \end{array}
   \right) \, ,
  ~~~~~~
   \sigma^\m = (1,\sigma^i) \, ,~~~~~
  \bar{\sigma}^\m = (-1,\sigma^i)\, ,~~~~~
  \g_5 \equiv i \g^0 \g^1 \g^2 \g^3 \, .
\eea
Note that $(\sigma^\m)_{\a\db} (\bar{\sigma}_\m)^{\gd \d} =  2 \, \delta_\a^{~\d} \, \delta_{\db}^{~\gd}$ and
$(\bar{\sigma}^\mu)^{\da \b}
  = - \eps^{\b \g} \eps^{\da \dd}  (\sigma^\mu)_{\g \dd}$ with
$\eps^{12} = \eps_{12} = 1$.

For null momenta $p_i$, the angle and square
spinors $|i\>$ and $|i]$ are defined as in \cite{Bianchi:2008pu}, and we follow the conventions of \cite{Bianchi:2008pu} for all
 two-component spinor brackets and their identities.

\subsection{{\em q}-helicity basis}
The massive Dirac equation is written
\bea
  \label{Dirac}
  (\slashed{p} - im)\, v_\pm(p) = 0 \, ,~~~~~~
  (\slashed{p} + im)\, u_\pm(p) = 0 \, ,~~~
\eea
and using $\bar{\psi} \equiv - i \psi^\dagger \g^0$ for the Dirac adjoint we have
\bea
  \label{bDirac}
  \overline{v}_\pm(p) \,(\slashed{p} - im)= 0 \, ,~~~~~~
  \overline{u}_\pm(p) \,(\slashed{p} + im) = 0 \, .~~~
\eea
We use the formalism of section \ref{s:massSH} to solve the Dirac equation.
The solutions are expressed in terms of 4-component spinor bra's and ket's:

\vspace{2mm}
\noindent {\bf Outgoing  anti-fermions:}
\bea
~~|p\dak ~\equiv~
  v_-(p) =
  \left(
  \begin{array}{c}
    \frac{im}{\a_p} |q]_\a \\[2mm]
    |p^\perp\>^{\db}
  \end{array}
  \right) ,~~~~~~~~
~~|p\dsk ~\equiv~
  v_+(p) =
  \left(
  \begin{array}{c}
    |p^\perp]_\a \\[2mm]
    \frac{im}{\a_p} |q\>^{\db}
  \end{array}
  \right) \,.
\eea
\noindent {\bf Outgoing  fermions:}
\bea
  \label{outferm}
  \dsb p| ~\equiv~
  -i\, \overline{u}_+
  = \Big( [p^\perp|^\a\, , \, -{\textstyle \frac{im}{\a_p} }\<q|_{\db} \Big)\,,~~~~~~~~
  \dab p| ~\equiv~
  i\, \overline{u}_-
  = \Big( -{\textstyle \frac{im}{\a_p} }[q|^{\a} \, ,\, \<p^\perp|_{\db}\,\Big)\,.~~~~~~~~
\eea

\vspace{2mm}
\noindent {\bf Incoming fermions:}
\bea
~~|p\dakB ~\equiv~
  u_+(p) =
  \left(
  \begin{array}{c}
    -\frac{im}{\a_p} |q]_\a \\[2mm]
    |p^\perp\>^{\db}
  \end{array}
  \right) ,~~~~~~~~
~~|p\dskB ~\equiv~
  u_-(p) =
  \left(
  \begin{array}{c}
    |p^\perp]_\a \\[2mm]
    -\frac{im}{\a_p} |q\>^{\db}
  \end{array}
  \right)\,.
\eea
\noindent {\bf Incoming anti-fermions:}
\bea
  \dsbB p| ~\equiv~
  -i\, \overline{v}_-
  = \Big( [p^\perp|^\a\, , \, {\textstyle \frac{im}{\a_p} }\<q|_{\db} \Big)\,,~~~~~~~~
  \dabB p| ~\equiv~
  i\, \overline{v}_+
  = \Big( {\textstyle \frac{im}{\a_p} }[q|^{\a} \, ,\, \<p^\perp|_{\db}\,\Big)\,.~~~~~~~~
\eea
In these expressions, we have introduced $ \a_p \equiv \sqrt{\<q|p|q]}$, which satisfies the following relations
\bea
  \label{qperp}
  \<q \,p^\perp \> ~=~ \frac{\<q|p|q]}{\alpha_p} ~=~ \a_p ~=~[ p^\perp\, q] \, .
\eea

To recover the familiar \emph{massless spinor-helicity formalism}, simply set $m=0$ and replace $p^\perp$ with the null momentum $p$.

It is easy to verify the orthogonality and normalization properties of the
 massive solutions directly:
$\overline{u}_s(p)  u_{s'}(p)  \,=\, 2 m\, \delta_{ss'}$,~
$\overline{v}_s(p) v_{s'}(p)  \,=\, -2 m\, \delta_{ss'}$, and
$\overline{u}_s(p) v_{s'}(p)  \,=\, 0 \,=\, \overline{v}_s(p) u_{s'}(p)$.
One needs $\<q \,p^\perp \> ~=~ \<q|p|q]/\alpha_p ~=~ \a_p ~=~[ p^\perp\, q]$.
These rules can be summarized in the bra-ket notation:
\bea
  \nonumber
  &&
  0 ~=~ \dsb p| p \dsk
  ~=~ \dab p| p \dak
  ~=~ \dab p| p \dsk
  ~=~ \dsb p| p \dak
  ~=~ \dsbB p| p \dskB
  ~=~ \dabB p| p \dakB
  ~=~ \dabB p| p \dskB
  ~=~ \dsbB p| p \dakB\,,\\[2mm]
  &&
  \nonumber
  0  ~=~\dabB p| p \dak ~=~  \dab p| p \dakB\, ~=~
  \dsbB p| p \dsk  ~=~ \dsb p| p \dskB\,, \\[2mm]
  &&
  \!\!\!\!\!\!\!2 i m~=~\dab p|p\dskB ~=~ \dsbB p | p \dak ~=~\!-\!\dabB p|p\dsk \!\!~=~ -\dsb p | p \dakB \,.
  \label{eq:doubleBracketIdentities}
\eea
The spinor completeness relations are given in the main text, see \reef{complete1}.

\vspace{2mm}
\noindent {\bf {\em q}-helicity:}
We can label the solutions of the massive Dirac equation by $q$-helicity $\tilde{h} = \pm\frac{1}{2}$. These states are eigenstates of the ``$q$-helicity'' operator
\bea
 \label{SigmaPM}
\tilde{\Sigma}^ \pm_{p;q} \equiv \frac{1}{2}(1 \pm i  \gamma_5 \slashed{\ell})\, ,
~~~~~~~~
\ell = \frac{1}{m}\big( p^\perp  - \frac{m^2}{\a_p^2} q \big) \,,
\eea
where $\ell$ is a unit vector orthogonal to $p$: $p \cdot \ell=0$ and $\ell^2=1$.
It is easy to see that
\bea
&&
    \tilde{\Sigma}^-_{p;q}  |p\dak = |p\dak \,,
    ~~~~~
    \tilde{\Sigma}^-_{p;q}  |p\dsk = 0\,,
    ~~~~~~~\!
    \tilde{\Sigma}^-_{p;q}  |p\dakB = 0 \,,
    ~~~~~~~~\!
    \tilde{\Sigma}^-_{p;q}  |p\dskB = |p\dskB \,, \\[2mm]
&&
    \tilde{\Sigma}^+_{p;q}  |p\dak = 0 \,,
    ~~~~~~~\,
    \tilde{\Sigma}^+_{p;q}  |p\dsk = |p\dsk \,,
    ~~~~
    \tilde{\Sigma}^+_{p;q}  |p\dakB = |p\dakB \,,
    ~~~~\,\,
    \tilde{\Sigma}^+_{p;q}  |p\dskB = 0\,.
\eea
Thus $|p\dak$ has $q$-helicity $\tilde{h}=-1/2$ while $|p\dsk$ has  $\tilde{h}=+1/2$. Similarly,  $|p\dakB$ has $q$-helicity $\tilde{h}=1/2$ while $|p\dskB$ has  $\tilde{h}=-1/2$.
In the massless limit, $q$-helicity reduces to the usual helicity label of massless states.

\vspace{2mm}
\noindent {\bf Polarizations of massive vector bosons:}
It is easy to verify that the polarization vectors defined in \reef{theEps} satisfy the orthonormality and completeness conditions
\bea
  p \cdot \eps_\l =0 \,,~~~~~~~~
  \eps_\l \cdot \eps_{\l'}^* = \delta_{\l\l'}\,, ~~~~~~~~
  \sum_{\l=+,0,-} {\eps_\l^\m}^* \eps_\l^\n = \eta^{\m\n} + \frac{p^\m p^\n}{m^2}
  \,.
\eea
Furthermore, since $\ell$ in the $q$-helicity operator \reef{SigmaPM} equals $\epsilon_0$, the polarization $\epsilon_0$ will be a longitudinal polarization with respect to the same spacelike direction $\ell$ as we use to label the Dirac spinor states.

\vspace{2mm}
\noindent {\bf Polarizations of massless vector bosons:}
Setting $m=0$ and replacing $p^\perp \to p$, we recover the well-known expressions
\bea
  \label{mzeroPol}
  \slashed{\eps}_-
  =
  \frac{\sqrt{2}}{[p\, q]}
  \Big(
     |p\> [q| -
    |q] \<p |
  \Big) \, , ~~~~~~~~~~
  \slashed{\eps}_+
  =
  \frac{\sqrt{2}}{\< p\, q \>}
  \Big(
    |q\> [p| -|p] \<q |
  \Big) \,
\eea
for the polarizations of  massless gauge bosons.

\vspace{2mm}
\noindent {\bf Crossing symmetry:}
Crossing symmetry takes $p_i \to -p_i$ and we also take $q_i \to -q_i$. The result is the crossing rules $|p\dakB \to |-p \dak$ and $|p\dskB \to |-p \dsk$ which are incorporated in the Feynman rules \reef{Dfeyn}.

\subsection{Helicity basis}
Above we have expressed the solutions to the Dirac equation in terms of the formalism used in section \ref{s:massSH}. For convenience, we record here the solutions in the more familiar helicity basis. We start by writing
\bea
   p^\m = (p^0,\,|\vec{p}|\, \vec{e})
   ~~~~~\text{with}~~~~~
   \vec{e} = (\cos\phi \sin\theta,\, \sin\phi \sin\theta,\, \cos\theta)\, .
\eea
Then
\bea
  p^{\da \b}~=~
\begin{pmatrix}
  p^0 + |\vec{p}|\, (c^2 -s^2) & |\vec{p}|\, e^{-i\phi}\, 2 c s \\[1mm]
  |\vec{p}|\, e^{i\phi}\, 2 c s & p^0 - |\vec{p}| \,(c^2 -s^2)
\end{pmatrix},
~~~~~~~
s \equiv \sin{\text{\footnotesize $\frac{\th}{2}$}}\,, ~~~~~
c \equiv \cos{\text{\footnotesize $\frac{\th}{2}$}}\,.
\eea
Define spinors $\kappa$ as solutions to the \emph{massive} coupled Weyl equations
\bea
  &&p^{\da\b}\, |\k_{1}]_{\b} = - m \,|\k_2\>^{\da} \, ,~~~~
    p^{\da\b}\, |\k_{2}]_{\b} = + m \,|\k_1\>^{\da}\, ,~~~~~~\\
  &&p_{\a\db}\, |\k_{1}\>^{\db} = -m \,|\k_{2}]_{\a} \, ,~~~~
    p_{\a\db}\, |\k_{2}\>^{\db} = + m \,|\k_{1}]_{\a}\, .~~~
\eea
For example $|\k_{1}]_{\a}= \sqrt{p^0  -|\vec{p}|}
  \begin{pmatrix}
   c\,  e^{-i\phi}  \\
   s
   \end{pmatrix}
$
and
 $|\k_{2}]_{\a}= \sqrt{p^0  +|\vec{p}|}
  \begin{pmatrix}
   s  \\
   -c\,  e^{i\phi}
   \end{pmatrix}
$.
The solutions to the massive Dirac equation \reef{Dirac} can then be written
 in terms of $\kappa_i$ as follows.

\vspace{1mm}
\noindent {\bf Outgoing:}
\bea
  |H^-\dak \equiv v_- =
 \begin{pmatrix}
    |\k_{1}]_{\a} \\[1mm]
    i\, |\k_{2}\>^{\da}
  \end{pmatrix} \, ,
  \hspace{5mm}
 &&
 |H^+\dsk \equiv v_+ =
 \begin{pmatrix}
    |\k_{2}]_{\a} \\[1mm]
    -i\, |\k_{1}\>^{\da}
  \end{pmatrix} \, ,
  \\[1mm]
\dsb H^+| \equiv - i\, \overline{u}_+ =
 \begin{pmatrix}
    i\,[\k_{2}|^{\a}\,,
      -\<\k_{1}|_{\da}
  \end{pmatrix} \, ,\,
  \hspace{5mm}
 &&\dab H^-| \equiv i\, \overline{u}_-=
 \begin{pmatrix}
    i\,[\k_{1}|^{\a}\,,
      \<\k_{2}|_{\da}
  \end{pmatrix}
  \, .
\eea

\vspace{1mm}
\noindent {\bf Incoming:}
\bea
 |H^+\dakB \equiv u_+ =
 \begin{pmatrix}
    |\k_{1}]_{\a} \\[1mm]
    -i\, |\k_{2}\>^{\da}
  \end{pmatrix} \, ,
  \hspace{5mm}
 &&
 |H^-\dskB \equiv u_- =
 \begin{pmatrix}
    |\k_{2}]_{\a} \\[1mm]
    i\, |\k_{1}\>^{\da}
  \end{pmatrix} \, ,
  \\[1mm]
  \hspace{5mm}
 \dsbB H^-| \equiv  - i\, \overline{v}_- =
 \begin{pmatrix}
    -i\,[\k_{2}|^{\a}\,,
      -\<\k_{1}|_{\da}
  \end{pmatrix} \, ,\,
 &&
\dabB H^+| \equiv i\, \overline{v}_+ =
 \begin{pmatrix}
    -i\,[\k_{1}|^{\a}\,,
       \<\k_{2}|_{\da}
  \end{pmatrix}
  \, .~~
\eea
The helicity projection operator is
\bea
  \Sigma^\pm_p = \frac{1}{2} (1 \pm i\, \g_5 \, \slashed{z}) \,.
\eea
In the rest frame $p^\m = (m,0,0,0)$ and $\Sigma_p^\pm$ projects along the $z$-axis, $z^\mu=(0,0,0,1)$. Boosting to a general frame, $z$ has to satisfy $z^\m z_\m = 1$ and $z\cdot p = 0$. If $p^\mu=(p^0,|\vec{p}|\, \vec{e})$ as above, then it is easy to see that
$z^\mu=\frac{1}{m} (|\vec{p}|,p^0 \,\vec{e})$ solves these conditions. One then finds
\bea
&&
\vspace{-2cm}
    \Sigma^-_{p;q}  |H_p^-\dak = |H_p^-\dak \,,
    ~~~~
   \Sigma^-_{p;q}  |H_p^+\dsk = 0\,,
    ~~~~~~~~\!
   {\Sigma}^-_{p;q}  |H_p^+\dakB = 0 \,,
    ~~~~~~~~~\!
   {\Sigma}^-_{p;q}  |H_p^-\dskB = |H_p^-\dskB \,, ~~~~~~~~~~ \\[2mm]\vspace{-2cm}
&&
    \Sigma^+_{p;q}  |H_p^-\dak = 0 \,,
    ~~~~~~~~~~
   {\Sigma}^+_{p;q}  |H_p^+\dsk = |H_p^+\dsk \,,
    ~~
   {\Sigma}^+_{p;q}  |H_p^+\dakB = |H_p^+\dakB \,,
    ~~
   {\Sigma}^+_{p;q}  |H_p^-\dskB = 0\,.
\eea
which justifies the (half-integer) helicity assignments $s=\pm$ on $H^s_p$.

Polarization for spin-1 particles can be expressed in this formalism as
\bea
  \epsilon_{-}^{\da\b} &=& \frac{\sqrt{2} |\k_2\>^{\da} [\k_1|^\b}{[\k_1 \k_2]}\, ,
  ~~~~~~
  \epsilon_{+}^{\da\b} ~=~ \frac{\sqrt{2} |\k_1\>^{\da} [\k_2|^\b}{ \<\k_1 \k_2\>} \, ,
  \\[2mm]
  \epsilon_L^{\da\b} \label{usualeps}
  &=& - \frac{1}{m} \big( |\k_{1}\>^{\da} [\k_{1}|^{\b} - |\k_{2}\>^{\da} [\k_{2}|^{\b} \big) \,.
  \hspace{8mm}
\eea
Plugging in the explicit solutions for $\k_{1,2}$ we find
\bea
  \eps^\mu_\pm(p)
   &=& \mp \frac{e^{\pm i\,\phi}}{\sqrt{2}}
     \Big( 0,~\cos\th \, \cos\phi \pm i \sin\phi, ~
      \cos\th \, \sin\phi \mp i \cos\phi , ~ - \sin\th \Big) \, ,\\[2mm]
  \eps^\mu_0(p)
  &=&
  \frac{p^0}{m} \Big(\,
  \frac{|\vec{p}|}{p^0}\,, ~\cos\phi \sin\theta\,,~ \sin\phi \sin\theta\,,~ \cos\theta\Big)
  \,.
\eea
This reduces to the familiar result in the restframe.

\subsection{Changing basis}
It is easy to switch from one basis to the other. For example, suppose we want to express the $q$-helicity state $|p\dak$ in the helicity basis. We just write
\bea
   |p\dak = a\, |H_p^-\dak + b\, |H_p^+\dsk \, .
\eea
Then use identities such as $\dabB H_p^- H_p^- \dak = 0$ and
$\dabB H_p^+ H_p^+ \dsk = -2i\, m$
to conclude that
\bea
   |p\dak  =
   \frac{1}{2i\,m}
   \Big(
        \dsbB H_p^- p \dak~ |H_p^-\dak -  \dabB H_p^+ p \dak~ |H_p^+\dsk
   \Big) \, .
\eea
Likewise polarizations in one basis can be obtained as a linearly combination of those in the other basis.

\end{appendix}

\small
\bibliographystyle{utphys}
\bibliography{AllLineShiftsAndCSW}

\end{document}